\documentclass[
 reprint,
showpacs,
preprintnumbers,
nobibnotes,
 amsmath,amssymb,
 aps,
prb,
floatfix
]{revtex4-2}

\usepackage{graphicx}
\usepackage{dcolumn}
\usepackage{bm}
\usepackage{amsmath,amssymb}
\usepackage{amsmath,mathtools}
\usepackage{leftindex}
\usepackage[colorlinks=true, linkcolor=blue, urlcolor=blue, citecolor=blue]{hyperref}

\newcommand{\qvec}[1]{\bm{#1}}

\begin{document}

\preprint{APS/123-QED}

\title{Analytic approach to creating homogeneous fields with finite-size
magnets}
\author{Ingo Rehberg}
\email{Ingo.Rehberg@uni-bayreuth.de}
\affiliation{Institute of Physics, University of Bayreuth, 95440 Bayreuth, Germany}
\author{Peter Blümler}
\email{bluemler@uni-mainz.de}
\affiliation{Institute of Physics, University of Mainz, 55128 Mainz, Germany}
\date{\today}

\begin{abstract}
Homogeneous magnetic fields can be generated through the strategic arrangement of permanent magnets. The Halbach array serves as a prominent example of an effective design following this principle. However, it is a two-dimensional approach because it is optimal when placing infinitely long magnets -- line dipoles-- on a circle. If shorter, more realistic magnets are to be used, the optimal arrangement of magnetic moments diverges from the classical Halbach geometry. This paper presents optimal solutions for three-dimensional arrangements calculated for point dipoles, including optimized orientations for single rings and stacks of two rings. They are superior to the original Halbach arrangement and a modification described in the literature, both in terms of the strength and the homogeneity of the magnetic field. Analytic formulae are provided for both cases and tested by experimental realizations.
\end{abstract}

\maketitle


\section{\label{sec:intro}Introduction}
\begin{figure}[ht]
\includegraphics[width=.48\textwidth]{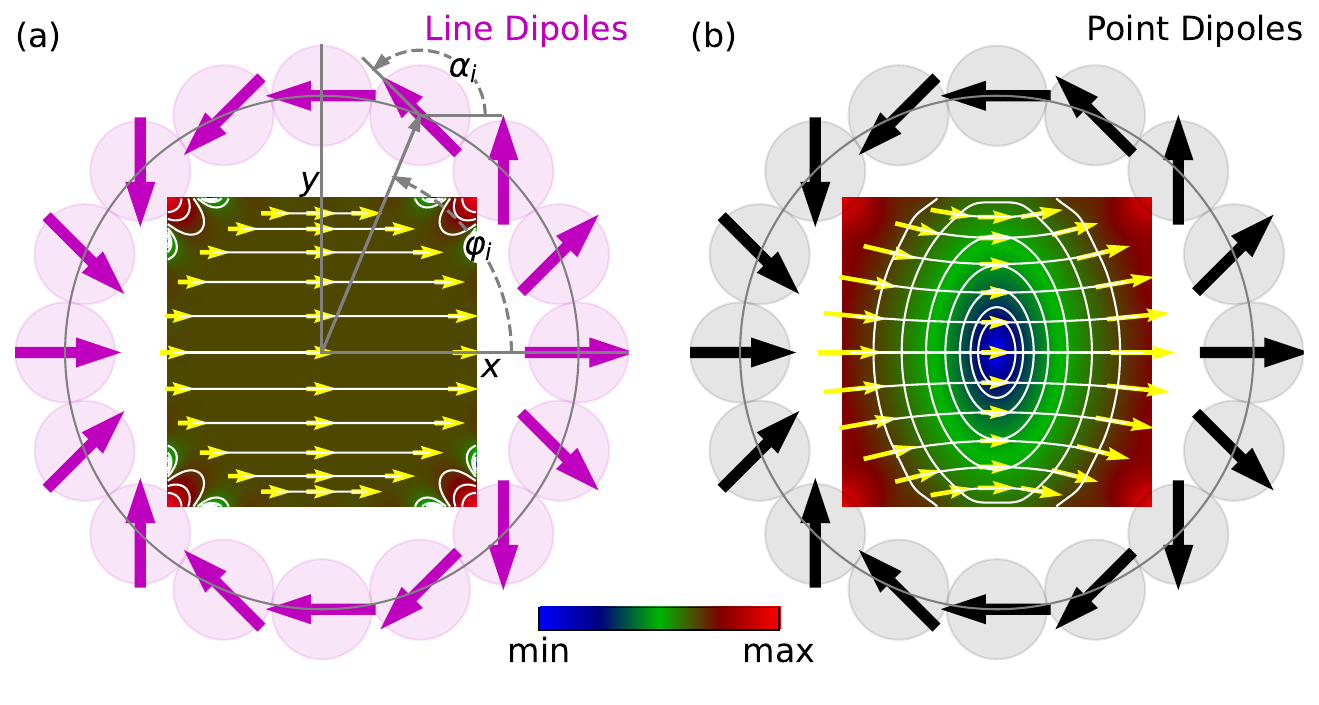}
\caption{Comparison of line and point dipoles. (a) Halbach arrangement of line dipoles (pink circles) with the direction of  magnetization  $\qvec{m'}$ (magenta arrows) given by $\alpha_i\!=\!2\varphi_i$. The central square shows  $B_\text{x}$(color bar). It is very homogeneous with all field vectors (yellow lines and arrows) pointing in the $x$-direction. (b) The same orientation of point dipoles (gray circles with black arrows) leads to an inhomogeneous field with a local minimum $B_\text{c}$ in the center. The white contour lines indicate $2^i\%$, i= 0...6, deviation from $B_\text{c}$.}
\label{fig:1}
\end{figure}

The generation of homogeneous magnetic fields by permanent magnets is of interest for both laboratory measurements and various technical applications~\cite{Coey2002, Raich2004, Coey_book, BluCasa2015, Blümler2021, Bakenecker2020, Soltner2023}. Applying the concept of one-sided fluxes~\cite{Mallinson1973}, Klaus Halbach~\cite{Halbach1980} has presented a perfect solution to this challenge which is illustrated in Fig.~\ref{fig:1}.

 Halbach considered two-dimensional magnets, aka line dipoles (see Appendix \ref{sec:appA}), which can be approximately realized by very long circular rods magnetized perpendicular to their axis~\cite{Coey2002, Coey_book}. The Halbach ring is formed by placing the cylinders in a circle with radius $R$, their cross section is shown in light magenta in Fig.~\ref{fig:1}(a).  Setting the orientation angle $\alpha$ of the magnetic moments indicated by the arrows to twice the location angle $\varphi$, a very homogeneous field ensues \footnote{With $N$ cylinders, the center forms a saddle point of order $N$. Field deviations grow $\propto \rho^N$ from there, as can be interactively explored~\cite{Rehberg2024, Rehberg2025}}, as shown in the inset of Fig.~\ref{fig:1}(a).
 
 Using the same arrangement with finite-size magnets introduces a problem. To illustrate this, we consider the field of homogeneously magnetized spheres, which can be perfectly modeled as point dipoles~\cite{Jackson1999}. The result is illustrated in Fig.~\ref{fig:1}(b), where the circles indicate the cross section of the spheres. The field now has a local minimum in the center, and the lines of equal field strength are roughly ellipsoidal, with the long axis of the ellipse oriented perpendicular to the magnetic field direction.  

Tewari~et~al.~\cite{Tewari2021} successfully addressed this issue. They numerically optimized the orientation angles $\alpha$ of the point dipoles to achieve the best homogeneity within a finite region of interest. That region was chosen as the area inside a circle of half the diameter of the dipole ring.

Here we present an analytical calculation for the orientation with the primary goal of maximizing the field strength in the center of the ring. The resulting configuration is shown to lead to a stronger field compared to both the circular Halbach~\cite{Halbach1980} and the homogeneity optimized arrangement~\cite{Tewari2021}. Moreover, it is also the most homogeneous one of these three arrangements,  where homogeneity is quantified by the curvature of the magnetic field strength at the center of the arrangement. 

This analytical approach is further refined by introducing an additional degree of freedom -- allowing the magnets to twist along the third dimension. This shifts the maximum of the homogeneous field out of the magnet plane and facilitates the construction of stacked ring assemblies (limited here to two) with improved field strength and homogeneity. Moreover, rotating the rings within such a stack further enhances the uniformity of the magnetic field (magnitude) within the magnet plane.

\section{\label{sec:theory}Theory}

We consider magnetic fields in a plane (cf.\ Fig.~\ref{fig:1}(a)).  The location of the $i^\text{th}$ magnet is given by $(R, \varphi_i, 0)$ in a cylindrical coordinate system $(\rho,\varphi,z)$ and its magnetic dipole moment $\qvec{m}_i \!=\! m (0,\alpha_i,0)$. The field of a point dipole $\qvec{m}_i$ at the center is then~\cite{Jackson1999}:
\begin{equation}\label{eq:B_dip}
\begin{split}
        \qvec{B}_i^{\text{point}}\,&=\, \frac{\mu_0}{4 \pi R^3} \Bigl(3\qvec{\hat{\rho}_i} (\qvec{m}_i \cdot \qvec{\hat{\rho}_i}) -\qvec{m}_i \Bigr),\\[2ex]
\text{with}\quad \qvec{\hat{\rho}_i} &= \begin{pmatrix}\cos\varphi_i\\ \sin\varphi_i\\z_i\\ \end{pmatrix}/\sqrt{1+z_i} \\
      \text{and} \quad \qvec{m}_i &= m \begin{pmatrix}\cos\alpha_i\\ \sin\alpha_i\\0\\ \end{pmatrix} \text{in Cartesian coordinates.}\\
\end{split}
\end{equation}
\subsection{\label{sec:theory_line}Line dipoles}

The original Halbach configuration is two-dimensional, assuming magnets of infinite length along the $z$-direction. For this case, one has to integrate over a continuum of point dipoles along this dimension to obtain the equation for such a line dipole~\cite{Coey_book}:
\begin{equation}\label{eq:2}
\qvec{B}_i^{\text{line}} =\int_{-\infty}^{+\infty}\!\! \qvec{B}_i^{\text{point}} \;\text{d}z =\frac{\mu_0}{4 \pi R^2} \Bigl( 4\qvec{\hat{\rho}_i} (\qvec{m'_i} \cdot \qvec{\hat{\rho}_i}) -2\qvec{m'_i}\Bigr).\\
\end{equation}
Here, the magnetic dipole moment $\qvec{m}$ is replaced by a dipole density $\qvec{m'}\!=\!\qvec{m}/l$,
a dipole moment per unit length~\cite{Coey_book}. An illustration of the differences between $\qvec{B}_i^{\text{point}}$ and $\qvec{B}_i^{\text{line}}$ is given in Appendix \ref{sec:appA}.

The Halbach condition for the best arrangement of the line dipoles can be derived by setting each $m'_i$ at an angle $\alpha_i$ that maximizes the field along the $x$-direction. For the derivation, the constant factors are irrelevant, it is sufficient to discuss the term describing the angular dependencies:
\begin{eqnarray}\label{eq:3}
\qvec{B}^{\text{line}} \,&=&\,\left( \begin{array}{c c} B_x \\[1ex] B_y\\[1ex] \end{array}\right) \\
 &\propto& 2
  \begin{pmatrix}\cos\varphi \\\sin\varphi\\ \end{pmatrix} \left[\begin{pmatrix}\cos\alpha \\\sin\alpha\\\end{pmatrix}\!\cdot\!\begin{pmatrix} \cos\varphi\\\sin\varphi\\\end{pmatrix} \right]- \begin{pmatrix}\cos\alpha \\\sin\alpha\\ \end{pmatrix}. \nonumber
\end{eqnarray}
The $x$-component of the magnetic field thus reads:
\begin{eqnarray}\label{eq:4}
\nonumber
B_x^{\text{line}}&\propto&2\cos^2\!\varphi\,\cos\alpha\! +\!2\cos\varphi\,\sin\varphi\,\sin\alpha\!-\!\cos\alpha\\
     &=&\cos(2\varphi)\cos\alpha +\sin(2\varphi)\sin\alpha.
\end{eqnarray}

To obtain the angle $\alpha$ for maximal  $B_x^{\text{line}}$,  Eq.~(\ref{eq:4}) is differentiated with respect to $\alpha$ and investigated for roots
\begin{equation}\label{eq:5}
\frac{\partial B_x^{\text{line}} }{\partial \alpha} = -\cos(2\varphi) \sin\alpha + \sin(2\varphi)\cos\alpha  = 0,
\end{equation}
resulting in
\begin{equation}\label{eq:6}
\tan\alpha = \tan(2\varphi).
\end{equation}
The root corresponding to a maximum is then
\begin{equation}\label{eq:alpha_H}
\alpha \,\equiv\,^{\text{H}}\!\alpha \,=\, 2\varphi.
\end{equation}

This condition produces the strongest and most homogeneous field, as already identified by Klaus Halbach in 1980 [1] within a more general framework. To differentiate this specific orientation angle, $\alpha$, along with all associated parameters, from other configurations discussed in this work, we designate it with the presuperscript "H". The corresponding arrangement and its field are shown in Fig.~\ref{fig:1}(a).

\subsection{\label{sec:comparison}Comparison of circular arrangements of point dipoles}

Line dipoles, i.e., two-dimensional, infinitely long magnets---realizable as extended circular rods magnetized perpendicular to their axis---generate very homogeneous fields when aligned according to Eq.~(\ref{eq:alpha_H}). In contrast, finite-length magnets inevitably introduce a larger level of inhomogeneity. As the elementary form of a finite-size magnet we consider point dipoles, which are experimentally realized by homogeneously magnetized spheres~\cite{Borgers2018, Hartung2018}.  If they are arranged in the orientation given by Eq.~(\ref{eq:alpha_H}), the field indicated in Fig.~\ref{fig:1}(b) ensues. It shows an anisotropic inhomogeneity, as the field along the $x$-axis drops faster than the one along the $y$-axis.

Although many magnet designs have successfully used the classical orientation $^{\text{H}}\!\alpha$ (see, e.~g.,~\cite{Raich2004,Stamenov2006,Soltner2010, Bjork2011,BluCasa2015,Schmidt2021,Soares2022,Soltner2023}), regardless of the use of finite-size magnets, Tewari et al.~\cite{Tewari2021} tackled this issue by numerically optimizing the field homogeneity for point dipoles inside a circle of radius $R/2$. In a first approach, they only allowed for variations in dipole orientation $\alpha$. They fitted their discrete results found for the individual dipoles $\alpha_i$ by a smooth curve \footnote{The original equation (6) in Ref.~\cite{Tewari2021} is $\alpha \!=\! 2\varphi + a \cos(2\varphi)$ because a different coordinate system was used (cf.\ their Fig.~1)}:
\begin{equation}\label{eq:8}
^{\text{h}}\!\alpha  \,=\, 2\varphi + a\, \sin(2\varphi) \qquad \text{with} \qquad a \approx - \frac{\pi}{8}.\\
\end{equation}
Because their optimization aimed to optimize the field homogeneity, we use an index "h" for their result.

\begin{figure}[ht]
\includegraphics[width=.483\textwidth]{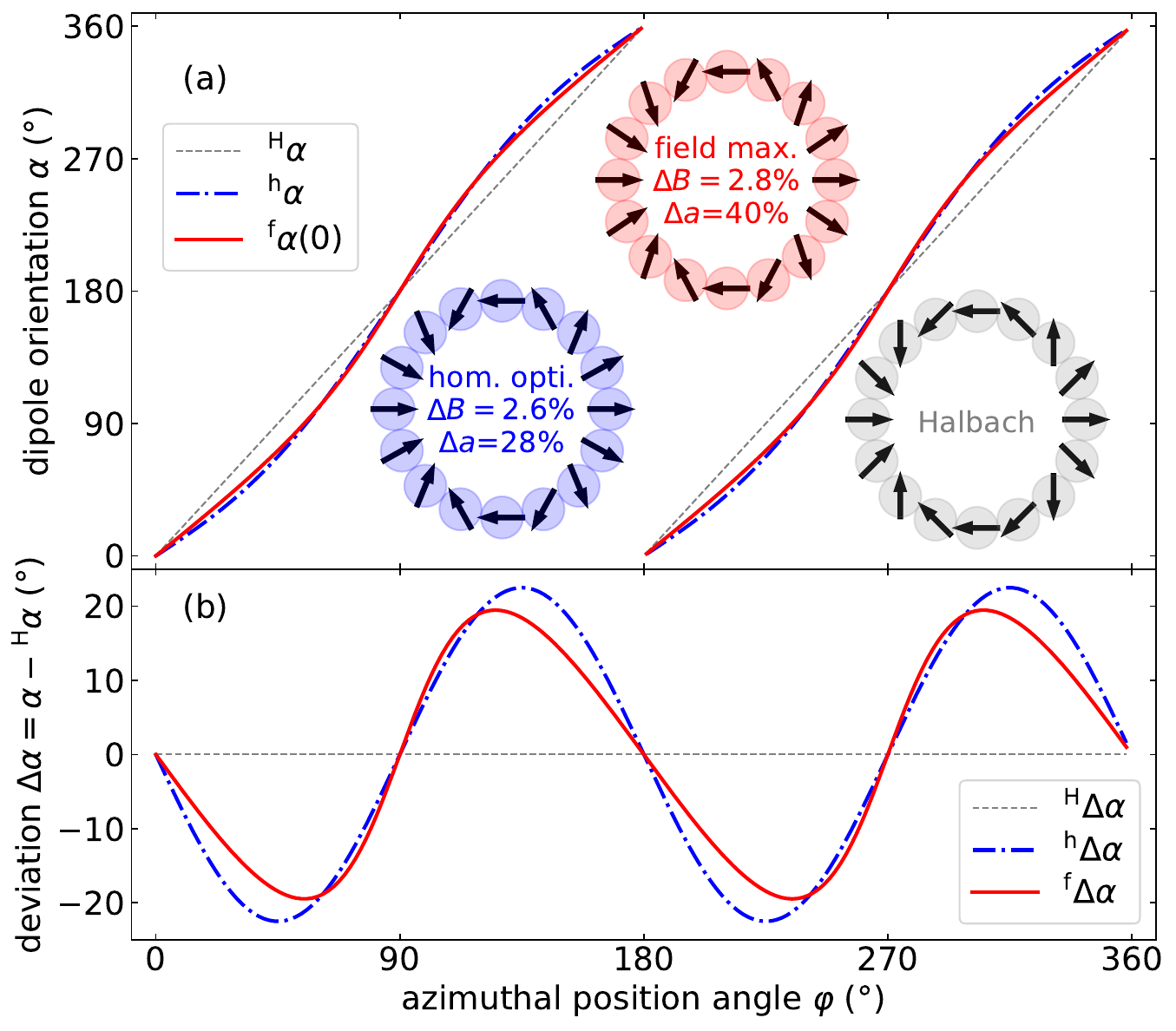}
\caption{Comparison of three different dipole arrangements. (a)  $^{\text{H}}\!\alpha$ (dashed black line, lower right gray sketch),  $^{\text{h}}\!\alpha$ (blue dashed-dotted line, lower left light blue sketch),  and $^{\text{f}}\!\alpha(0)$ (red line, upper center red sketch). (b) Enhancing the differences between the three configurations by subtracting $^{\text{H}}\!\alpha$.}
\label{fig:2}
\end{figure}

In this paper, we optimize the orientation for the strongest field generated by point dipoles by repeating the procedure for calculating $^{\text{H}}\!\alpha$. Starting from Eq.~(\ref{eq:B_dip}) with $z_i\!=\!0$, the analog to Eq.~(\ref{eq:4}) looks like: 
\begin{equation}\label{eq:9}
B_x^{\text{point}}\propto (3\cos^2\!\varphi - 1)\cos\alpha +\tfrac{3}{2} \sin(2\varphi)\sin\alpha.
\end{equation}
The maximal field is found analogously to Eq.~(\ref{eq:5}):
\begin{equation}
\frac{\partial B_x^{\text{point}}}{\partial \alpha}\propto(1-3\cos^2\!\varphi)\sin\alpha+\tfrac{3}{2} \sin(2\varphi)\cos\alpha.
\end{equation}
Setting this term to zero leads to
\begin{equation}
\tan\alpha=\frac{3 \sin(2 \varphi)}{6 \cos^2\!\varphi - 2},
\end{equation}
which is solved by 
\begin{equation}\label{eq:alpha_B}
 \qquad \alpha   \equiv \,
^{\text{f}}\!\alpha (0) = \tan^{-1} \left( \frac{3 \sin(2 \varphi)}{6 \cos^2\!\varphi - 2} \right) + c\,\pi, \,\, c \in \mathbb{Z}.
\end{equation}

The notation as $^\text{f}\!\alpha (0)$ refers to a special case of a more general description, which will be introduced in the following section. Eq.~(\ref{eq:alpha_B}) represents both the minima and the maxima of $B_\text{x}$. The angle \(\alpha\), with the correct choice of \(c\) to obtain a maximum (cf.~Appendix~\ref{sec:appB_new}), is most conveniently computed using the standard two-argument inverse tangent function \texttt{atan2(\(n, d\))}~\cite{Iso9899, Press1992}, which returns the principal value \(\alpha \in (-\pi, \pi]\), based on the signs of the numerator \(n\) and the denominator \(d\) \footnote{The unique value for $\alpha_\text{f}$ in programming notation reads
$\texttt{mod(atan2(3*sin(2*phi),6*cos(phi)}\wedge\texttt{2-2),2*pi)}$}.

The layout of these three dipole arrangements ($^{\text{H}}\!\alpha$, $^{\text{h}}\!\alpha$, and $^{\text{f}}\!\alpha(0)$) is presented in Fig.~\ref{fig:2}, showcasing the orientation of 16 dipoles as an example. The numbers for $\Delta b$ and $\Delta a$ within the sketched rings characterize their field strength and homogeneity in comparison to the Halbach configuration, as will be further quantified in connection with Fig.~\ref{fig:4}. The orientation of the dipoles located along $\varphi$ is quantitatively given by the smooth curves for $\alpha$.  Both $^{\text{h}}\!\alpha$ and $^{\text{f}}\!\alpha(0)$ deviate from the straight line representing $^{\text{H}}\!\alpha$. These differences and the small difference between $^{\text{h}}\!\alpha$ and $^{\text{f}}\!\alpha(0)$ appear more clearly when subtracting $^{\text{H}}\!\alpha$ from both functions, as shown in Fig.~\ref{fig:2}(b).

\begin{figure}[ht]
\includegraphics[width=.48\textwidth]{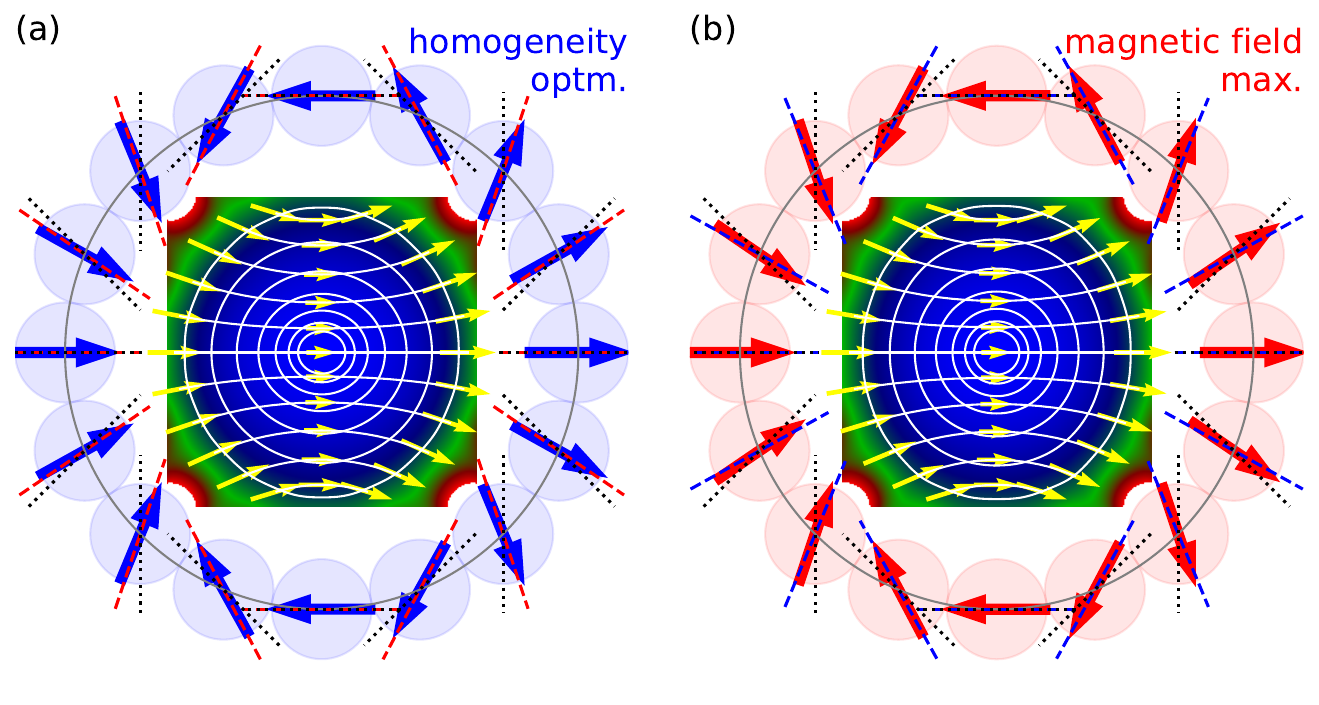}
\caption{Comparison of the two optimized configurations. (a)~$^{\text{h}}\!\alpha$ (blue arrows) , $^{\text{f}}\!\alpha(0)$ (red dashed),   and $^{\text{H}}\!\alpha$ (black dotted). Insets similar to Fig.~\ref{fig:1}(b). (b)~$^{\text{f}}\!\alpha(0)$ (red arrows) , $^{\text{h}}\!\alpha$ (blue dashed),   and $^{\text{H}}\!\alpha$ (black dotted).}
\label{fig:3}
\end{figure}

\begin{figure}[ht]
\includegraphics[width=.48\textwidth]{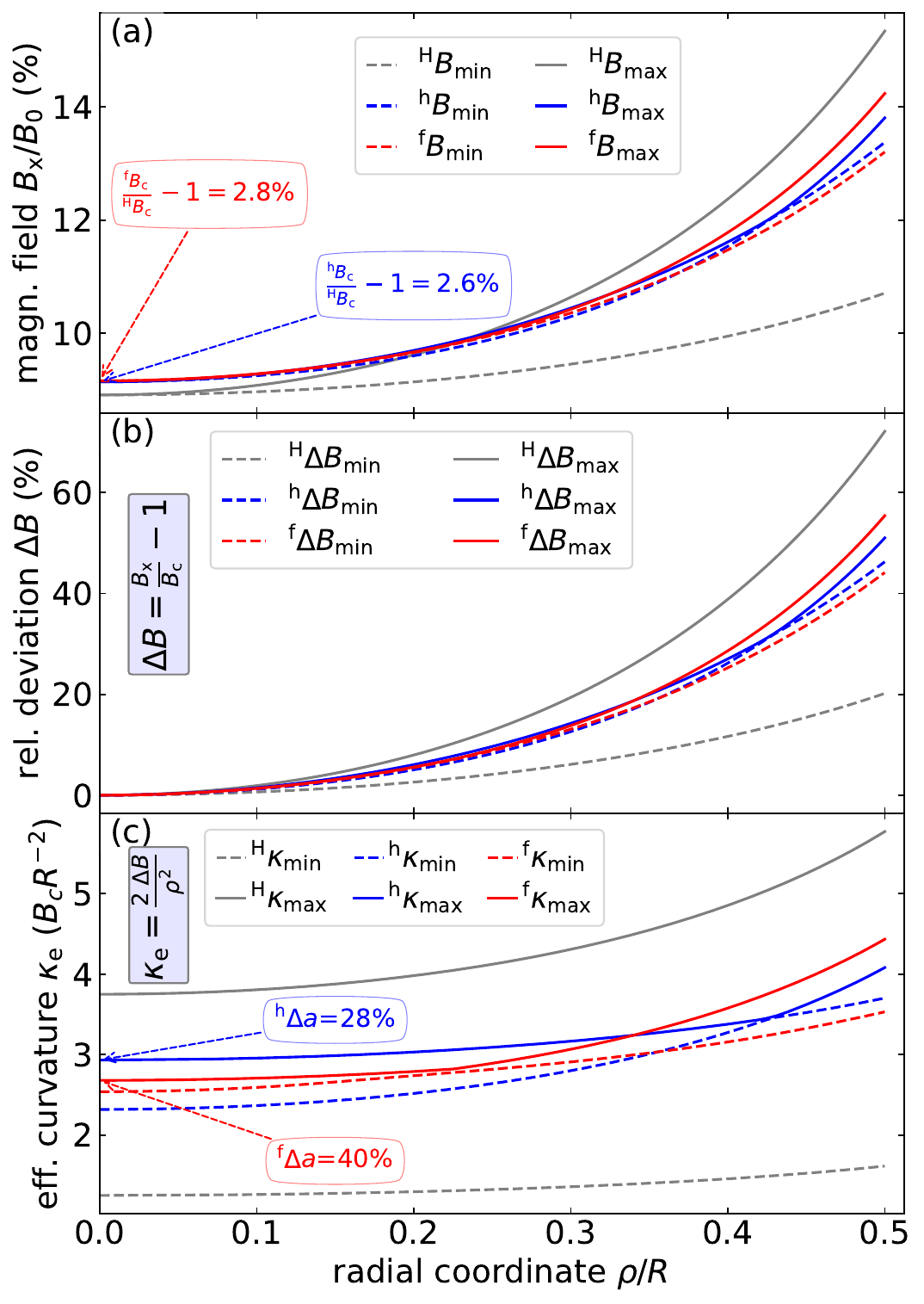}
\caption{Quantitative comparison of three configurations. (a) The minima (maxima) on a circle with radius $\rho$ are shown as dashed (solid) lines. The increase of the field compared to the Halbach configuration, $\Delta f$,  is given in the legend. (b) The deviation of the minimal and maximal fields from the value at the center, $B_\text{c}$, in units of $B_\text{c}$. (c) The effective curvature obtained from the minimum and maximum curves shown in (b). The corresponding relative area changes $\Delta a$ are given in the legend.}
\label{fig:4}
\end{figure}

To further highlight the subtile differences between the field-optimized and homogeneity-optimized configurations, both orientations are shown in Fig.~\ref{fig:3}. The resulting field in the plane is shown in the central insets. The contour lines of equal field strength $B_\text{x}$ are clearly rounder compared to the elliptical ones shown in Fig.~\ref{fig:1}(b), which are generated by the Halbach configuration of point dipoles. The difference in fields obtained by the $^{\text{h}}\!\alpha$- and the $^{\text{f}}\!\alpha(0)$-configuration needs a more quantitative investigation, as provided in Fig.~\ref{fig:4}.

In Fig.~\ref{fig:4}(a), the value of $B_x(x,y,0)$ has been calculated along a circle of radius $\rho\!=\!\sqrt{x^2+y^2}$. The field is scaled by
\begin{equation}\label{B_0}
B_0\!=\!\frac{2\mu_0m}{4\pi r_\text{s}^3}, 
\end{equation}
where $r_\text{s}$ is half the distance between the neighboring dipoles on the ring. For magnetic spheres of radius $r_\text{s}$ (which are indicated in contact in Figs.~\ref{fig:1}(b) and \ref{fig:3}),  $B_0$ is the field at the magnetic north pole.  This is the maximum field which can be measured at the surface of the sphere, and as such, it is a convenient scaling number. As a rule of thumb for neodymium magnetic spheres, one might think in terms of $100 \,\text{mT}$ here~\cite{Borgers2018, Hartung2018}, independent of the radius of the sphere. 

The minimum and maximum values of $B_x(x,y,0)$ are plotted as a function of $\rho$. It can be clearly seen that both values, as well as the gap between them, increase with increasing radius $\rho$ for all three configurations. It is also obvious that the center fields of the $^{\text{h}}\!\alpha$- and $^{\text{f}}\!\alpha(0)$-configuration are slightly larger than that of the Halbach arrangement ($^\text{H}\!\alpha$). The relative change is provided in the caption.  The center field in the $^{\text{f}}\!\alpha(0)$-configuration is the largest -- reflecting its deliberate design. However, the 2\textperthousand\  increase compared to the
$^{\text{h}}\!\alpha$-configuration is likely below the experimental resolution in many cases.

To compare the relative homogeneity of the three configurations, it is convenient to scale the fields with their value at the center,  $B_\text{c}$. The relative deviation from the center field $\Delta B \!=\! B_\text{x}/B_\text{c}-1$ is shown in Fig.~\ref{fig:4}(b). To give an example: If one is willing to accept a 10\%-deviation from the center field, the Halbach configuration could be used within a circle of $\rho/R \approx 0.22$, for the $^{\text{h}}\!\alpha$-configuration one gets $\rho/R \approx 0.25$, and for the $^{\text{f}}\!\alpha(0)$-configuration $\rho/R\approx 0.27$. Note that for a radius  $\rho/R \!=\! 0.5$ the $^{\text{h}}\!\alpha$-configuration shows the smallest deviation. That is to be expected, because the configuration was designed to minimize the deviation within that radius~\cite{Tewari2021}. Whether the corresponding 51~\% deviation is tolerable will ultimately depend on the specific application.

If one wants to compare the homogeneity of the different configurations independently from some arbitrarily chosen distance from the center, the curvature of the field strength around the center is a useful number. It could be obtained by taking the second derivative $\text{d}^2 \Delta B^\mathrm{max}(\rho)/\text{d}\rho^2$ at $\rho\!=\!0$. A practical approach is the calculation of the effective curvature 
\begin{equation}\label{kappa}
\kappa_\text{e}(\rho) =\frac{2 \,\, \Delta B^\mathrm{max}(\rho)}{\rho^2},
\end{equation}
which coincides with the curvature at $\rho\!=\!0$, and also emphasizes the deviations from the local parabola around the center.

The corresponding values are shown in Fig.~\ref{fig:4}(c).
The effective curvature obtained from $\Delta B ^\text{max}$ is inversely proportional to the radius of the homogeneous area near the center.  The increase of that area compared to the Halbach arrangement is obtained as $^{\text{f}}\!\Delta a \!=\!\, ^{\text{H}}\!\kappa(0)/^{\text{f}}\!\kappa(0)$, or $^{\text{h}}\!\Delta a \!=\!\, ^{\text{H}}\!\kappa(0)/^{\text{h}}\!\kappa(0)$ respectively. These ratios are independent of the specific deviation one might be willing to tolerate, assuming one stays in a regime where the effective curvature can be considered as constant. That is approximately the case for $\rho \lesssim 0.15 R$, where the deviations in $\Delta B$ are on a percentage level.

The plot clearly shows that the $^{\text{f}}\!\alpha(0)$-configuration offers the best homogeneity in a local area around the center. However, for regions with a radius greater than $0.34\;R$, the $^{\text{h}}\!\alpha$-configuration performs better, although at that level we are already dealing with a deviation of $\Delta B \approx 20\%$.

In conclusion of these theoretical considerations for the plane inside a dipole ring, the field-optimized $^{\text{f}}\!\alpha(0)$-configuration also shows the best local homogeneity. This underscores the desirability of experimentally realizing this particular configuration, as will be described in Section~\ref{sec:experiment}.

\subsection{\label{sec:finite_dist}Comparing arrangements of point dipoles at a finite distance from the ring}

Despite the optimizations described so far, it is obvious that an ideal line dipole would still provide the best solution to achieve a homogeneous magnetic field distribution inside the ring. A practical step toward this configuration is the construction of a stack of rings made from point dipoles. The working plane is then no longer located in the plane of a ring, but rather at a height $h\!=\!z/R$. Consequently, we set $z_i\!=\!h$ in Eq.~(\ref{eq:B_dip}). Moreover, we allow for a finite tilting angle $\theta$ of the dipoles with respect to the plane of the ring, i.\ e., 
\begin{equation}\label{eq:dip_mom}
\qquad \qvec{m} = m \begin{pmatrix}\cos\alpha\,\cos\theta\\ \sin\alpha\,\cos\theta\\\sin\theta\\ \end{pmatrix}.
\end{equation}
The analog to Eq.~(\ref{eq:4}) then reads
\begin{equation}\label{eq:B(h)}
\begin{split}
B_x^{\text{point}}\propto 
\frac{3\cos\varphi [\cos(\varphi\!-\!\alpha)\cos\theta\!+\!h\sin\theta]\!-\!(1\!+\!h^2)\cos\alpha}{(1+h^2)^{\frac{5}{2}}}.
\end{split}
\end{equation}

$B_x$ is now optimized for a specific value of $h$, referred to as $f\!=\!z/R$ in the following, because it can be interpreted as the focal length of the magnet arrangement. Setting the derivative of (\ref{eq:B(h)}) to zero leads to

\begin{equation}\label{eq:alpha_f}
^{\text{f}}\!\alpha(f) = \tan^{-1} \left( \frac{3 \sin(2 \varphi)}{6 \cos^2\!\varphi - 2 - 2f^2} \right). 
\end{equation}
It is remarkable that this result for the azimuthal angle $\alpha$ does not depend on the value of the tilt angle $\theta$.

\begin{figure}[ht]
\includegraphics[width=.48\textwidth]{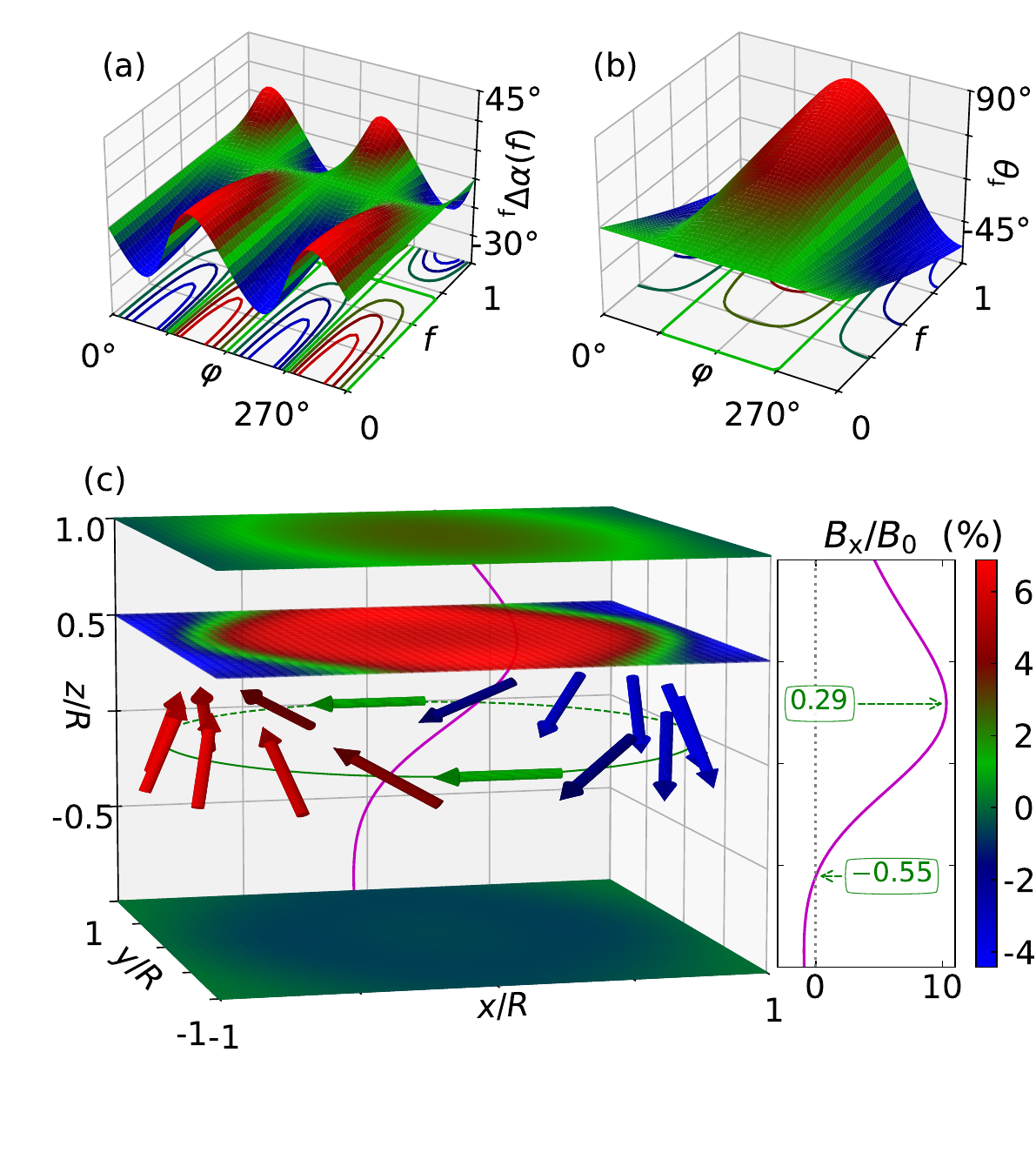}
\caption{Illustration of the $ ^{\text{f}}\alpha(f)$-configuration. (a) Dependence of $^{\text{f}}\!\Delta\alpha(f) $ on $\varphi$ and $f$ shown in a surface on top of a contour plot. The green contour line represents $^{\text{f}}\!\Delta\alpha(f)\!=\!0$. (b) Same presentation for the tilt angle $^\text{f}\theta$. (c) 16 dipoles, located on a ring at $h\!=\!0$, focusing to $f\!=\!1$. Arrows mark the direction of the moments (red :upwards, green: horizontal, blue: downwards). The strength of $B_\text{x}(x,y,h)$  is color-coded in the planes at $h\!=\!-1,1/2, \text{ and }1$. The central field $B_x(0,0,h)/B_0$ plotted next to it (magenta line) has a maximum at $h\!=\!0.29$. It is also projected onto the background plane $y\!=\!-R$ (arb. units).}
\label{fig:5}
\end{figure}

Equation (\ref{eq:alpha_f}) is illustrated in Fig.~\ref{fig:5}(a). The difference $^{\text{f}}\!\Delta\alpha(f) \!= ^{\text{f}}\!\alpha(f) -^{\text{H}}\!\alpha$ is shown to enhance the contrast, similar to  Fig.~\ref{fig:2}(b), where the line for $f\!=\!0$ is the one shown as $^{\text{f}}\!\Delta\alpha(0)$. Increasing $f$ from zero leads to a smaller variation of $^{\text{f}}\!\Delta\alpha(f)$ along the ring. At $f\!=\!1/\sqrt{2}$, the deviation is exactly zero. For farther distances, the deviation increases again, but with a change of sign for a fixed location $\varphi$ on the ring. 

Optimizing $B_\text{x}$ with respect to the tilt angle $\theta$ leads to
\begin{equation}\label{eq:theta_f}
^\text{f}\theta(f) = \tan^{-1} \left( \frac{-6 f\cos\varphi}{3 \cos(^{\text{f}}\!\Delta\alpha(f)) +(1-2f^2)\cos \left(^{\text{f}}\!\alpha(f)\right)} \right).
\end{equation}
This tilt angle $^\text{f}\theta(f)$ is illustrated in Fig.~\ref{fig:5}(b). It is zero at all angular positions $\varphi$ when $f\!=\!0$. For the angular positions $\varphi \!=\! 90^{\circ}$ and $\varphi \!=\! 270^{\circ}$ it is also zero for any focal length $f$, as indicated by the green contour lines in Fig.~\ref{fig:5}(b). 

A dipole arrangement that focuses on a finite distance $f$ by fulfilling both Eq.~(\ref{eq:alpha_f}) and Eq.~(\ref{eq:theta_f})  will be denoted as the $^{\text{f}}\!\alpha(f)$-configuration. For $f\!=\!0$, it coincides with the dipole angles $^{\text{f}}\!\alpha(0)$, which justifies the prior introduction in Eq.~(\ref{eq:alpha_B}). An example with the specific value $f\!=\!1$ is given in Fig.~\ref{fig:5}(c). 

The resulting $B_\text{x}$-field is shown in the $xy$-planes at $h\!=\!-1$,  $h\!=\!1/2$ and $h\!=\!1$, illustrating the fact that the $^{\text{f}}\!\alpha(+1)$ arrangement favors the positive $h$ direction. This is shown quantitatively by $B_x(0,0,h)/B_0$ (magenta line in Fig.~\ref{fig:5}(c)). It reveals a zero crossing at $h\approx -0.55$ and small fields below that distance from the ring.  The maximum of the field at $h\approx 0.29$ above the ring is approximately ten times larger, reaching a value of about $0.1B_0$.

\begin{figure}[ht]
\includegraphics[width=.48\textwidth]{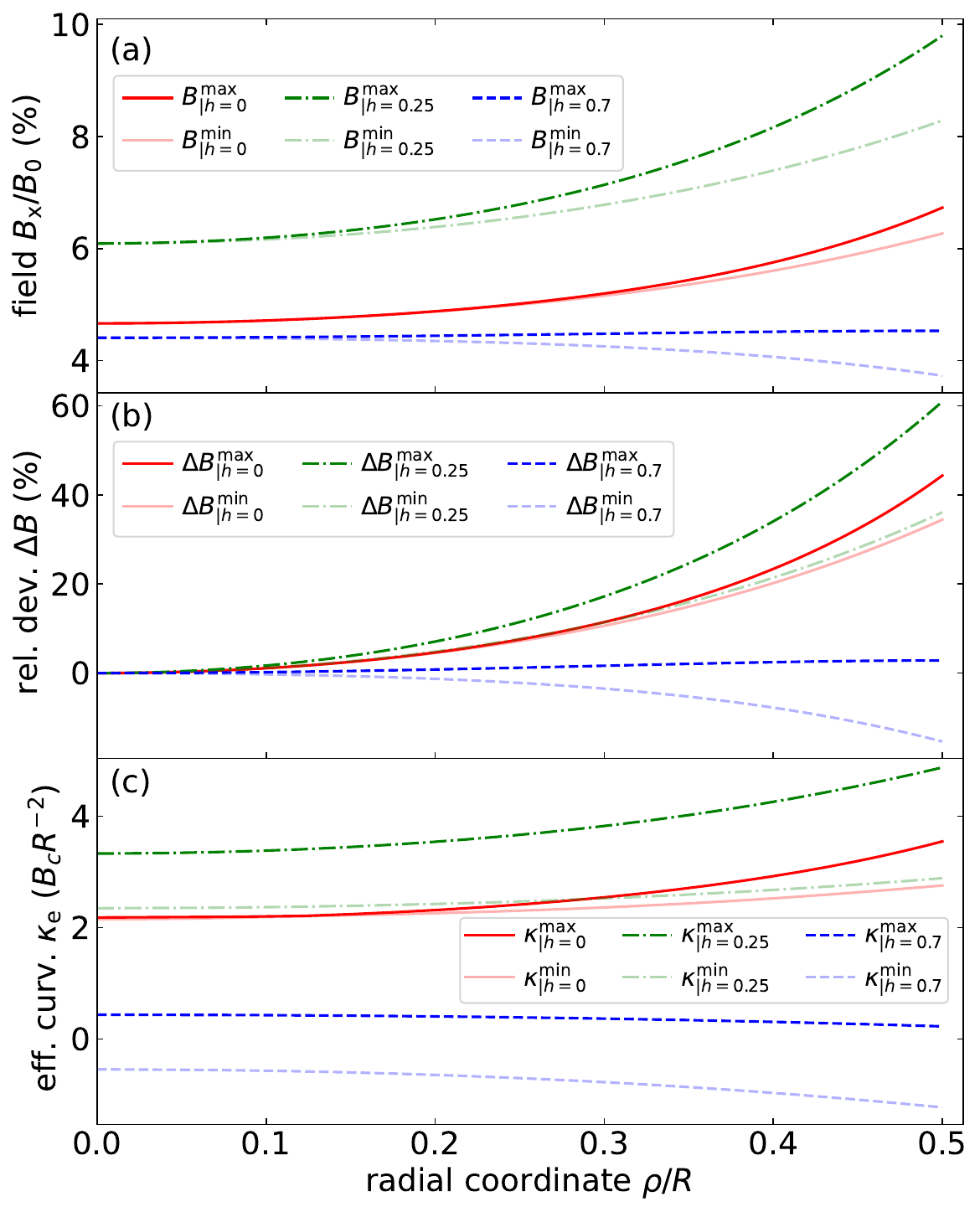}
\caption{Characterization of the $^{\text{f}}\!\alpha(1)$-arrangement by its field in three planes $h\!=\!0$, $0.25$, and $0.7$. The nomenclature is the same as in Fig.~\ref{fig:4}.  (a) For $h\!=\!0.25$, the magnetic field is larger than in the other two planes. (b) For $h\!=\!0.7$, the best homogeneity is achieved. Here the maximum shows a small positive curvature, and the minimum a negative one. (c)~The curvatures $\kappa$ as derived from $\Delta B$.}
\label{fig:6}
\end{figure}

 Figure\,\ref{fig:6} characterizes the arrangement of Fig.~\ref{fig:5}(c) with $^\text{f}\alpha(1)$ and the corresponding $^\text{f}\theta(1)$ in more detail. Fig.~\ref{fig:6}(a) shows the fields $B_\text{x}(x, y, 0R),$ $B_\text{x}(x, y, R/4)$, and $B_\text{x}(x, y, 0.7R)$, all scaled by $B_0$ as in Fig.~\ref{fig:4}. The field in the plane with $h\!=\!0$, where the dipoles are located, is smaller than the one shown in Fig.~\ref{fig:4} because the dipoles are not oriented in that plane. However, it is larger for $h\!=\!1/4$, a striking feature of the focusing property of this type of dipole ring. At distance $h\!=\!0.7$, the field strength is slightly lower than at $h\!=\!0$.

Figure~\ref{fig:6}(b) shows the relative deviations of the $B_\text{x}$ field components in the three planes by comparing them with the value in the center, similar to Fig.~\ref{fig:4}(b). The homogeneity is best at distance $h\!=\!0.7$. Here, the curvature for the maximal deviations is positive, and the one for the minimal values of $\Delta B$ is negative, that is, $B_\text{x}$ forms a saddle point in the center. Fig.~\ref{fig:6}(c) illustrates the two curvatures, minimum and maximum, for each of the three distances. The plane at a distance $h\!=\!0.7$ is notable for its excellent homogeneity. However, the field is reduced by a factor of about 0.55, compared to the "in-plane" arrangement discussed in Fig.~\ref{fig:4}.

\begin{figure}[ht]
\includegraphics[width=.48\textwidth]{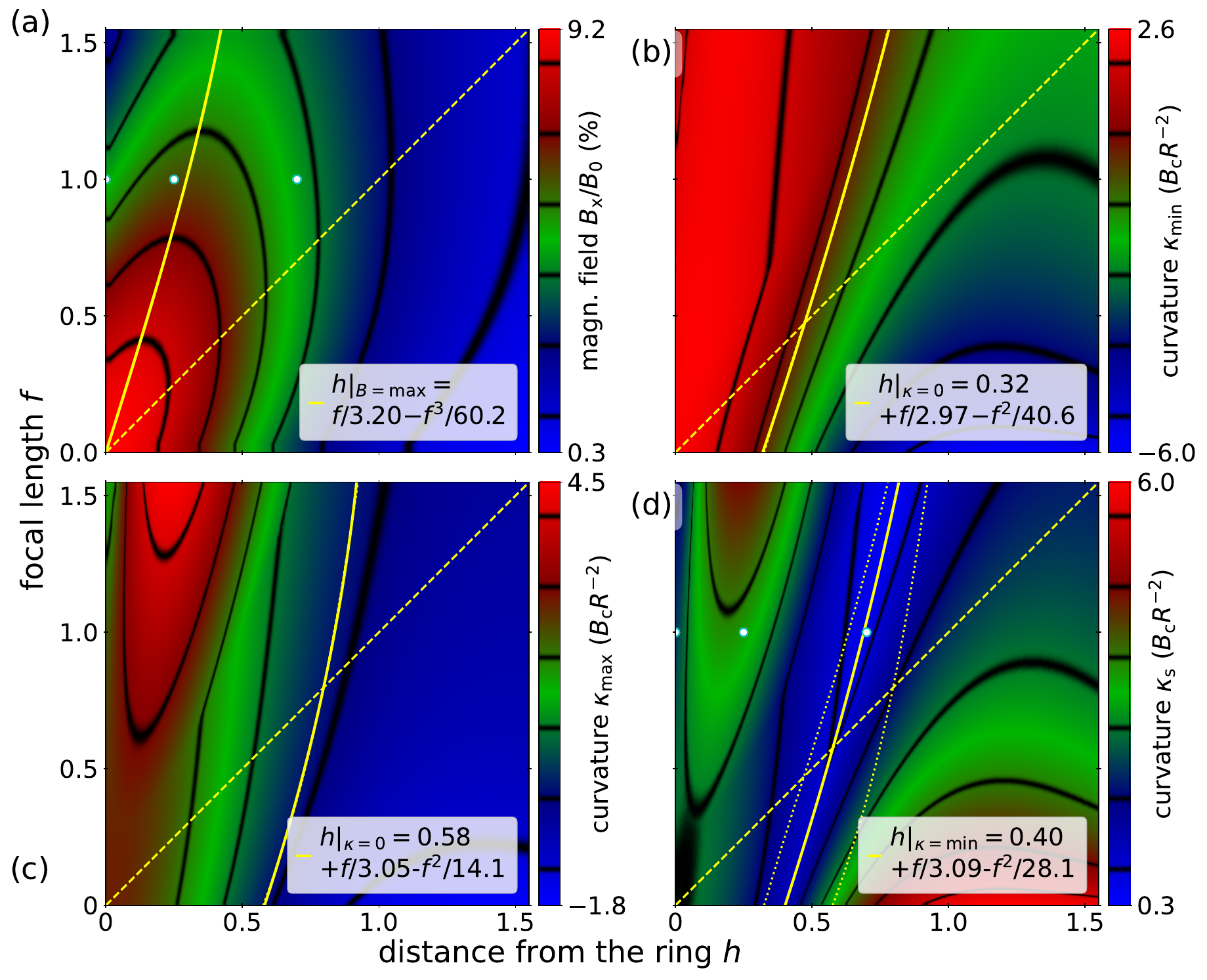}
\caption{Characterization of the focused arrangement. (a)\,The field strength $B_x(0,0,h)$ is color coded. The height for the field maximum is about $f/3$  (yellow line, a polynomial fit with the fitting values given in the legend. The same holds for (b),(c), and (d)).  The position of the focus (dashed yellow line) is shown in all four plots. The three heights discussed in Fig.~\ref{fig:6} are marked by cyan-white dots. (b,\,c)~The curvatures $\kappa_\text{min}$ and $\kappa_\text{max}$  with their zero crossings (yellow lines).  (d)\,The salient curvature $\kappa_\text{s}$. Its minimum (yellow line) is located between the zero crossings of $\kappa_\text{min}$ and $\kappa_\text{max}$ (dotted yellow lines) and crosses the focus (dashed yellow line) at $h\approx 0.576$.}
\label{fig:7}
\end{figure}

The field $B_x(0,0,h R)$ is shown in Fig.~\ref{fig:7}(a) as a contour plot with two control parameters: The distance from the ring $h$, and the focal length $f$. The field has a maximum of about $0.092 B_0$ in the lower left corner. That is, the center of the ring, with a focal length adjustment of $f\!=\!0$ -- which corresponds to $^\text{f}\theta(0)\!=\!0$ for the dipoles on the ring. This center field of that configuration decreases monotonically with the distance from the plane $h$. This changes qualitatively for finite values of $f$: The field increases from the center and reaches a maximum at about
\begin{equation}\label{eq:h_max}
h|_{B=\text{max}} \approx \frac{f}{3.20}-\frac{f^3}{60.2}, 
\end{equation}
and decreases monotonically from there. The location of that maximum is indicated by the solid yellow line. On the other hand, to maximize the field at a given distance $h$, one has to set the focal length $f$ of the dipole ring to this height $h$. The corresponding line $f\!=\!h$ is marked for guidance as a dashed yellow line in all four plots.

The homogeneity of the $B_x$-field can be characterized using a curvature $\kappa$. To be precise, two curvature values are determined: $\kappa_\text{max}$, corresponding to the maximum, and $\kappa_\text{min}$, corresponding to the minimum, for a circular trajectory with radius $\rho$ centered at the origin. As shown in Fig.~\ref{fig:6}, both values can be positive or negative. As the salient value describing the homogeneity of the field, we take the maximum of the absolute values of both numbers,
\begin{equation}\label{eq:kappa_s}
\kappa_\text{s}= \max(|\kappa_\text{min} |,| \kappa_\text{max}|).
\end{equation}
This curvature $\kappa_\text{s}$ is shown in Fig.~\ref{fig:7}(d) as a contour plot. It is remarkable that even for $f\!=\!0$ the location with the best homogeneity of the field lies at a final distance $h\!\approx\!0.40$.
\begin{figure}[ht]
\includegraphics[width=.48\textwidth]{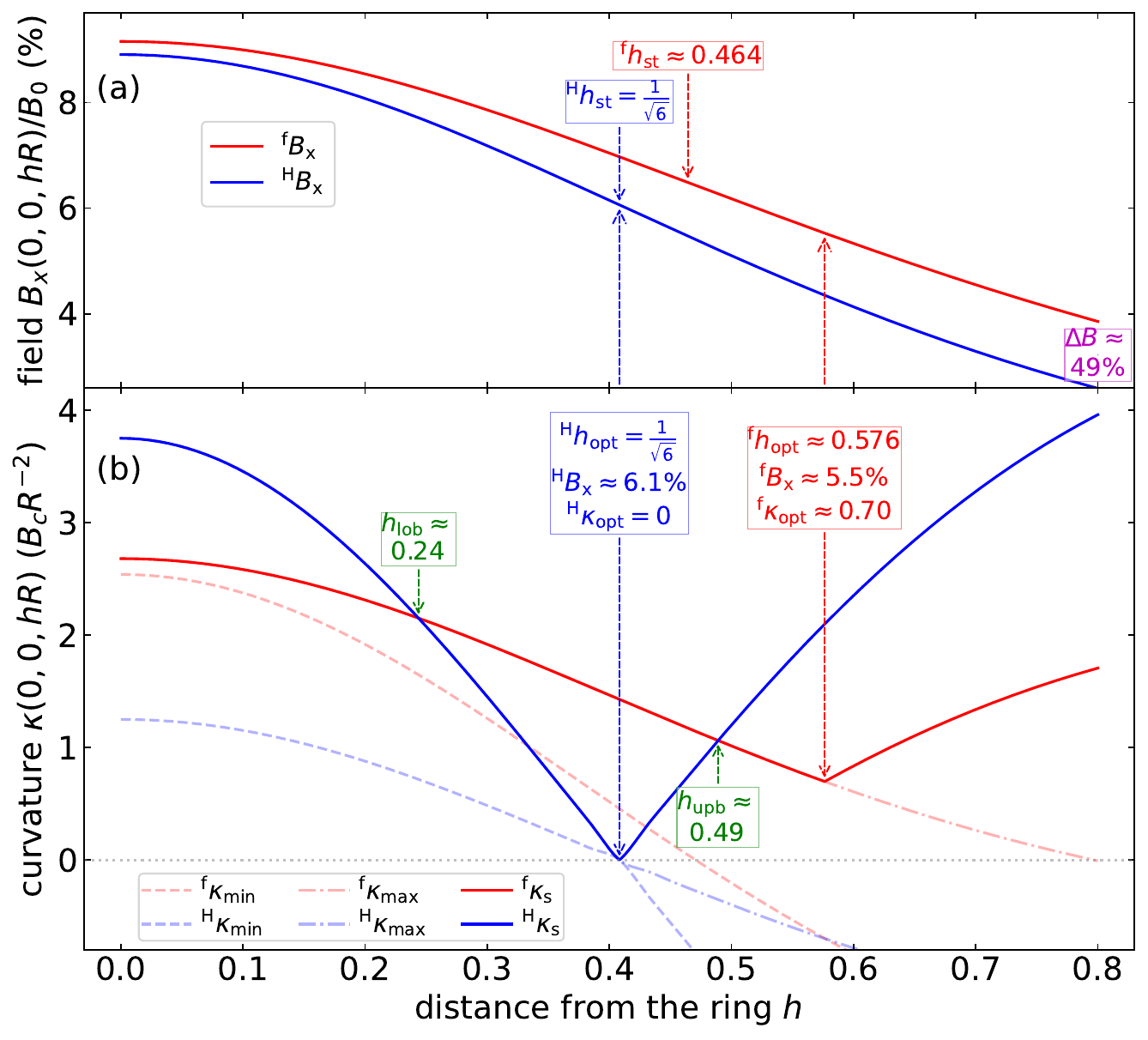}
\caption{Comparing the Halbach configuration (blue) with the configuration focused to $f\!=\!h$ (red). (a)\,Both fields decay monotonically. Their point of steepest descent, $h_\text{st}$, is annotated. The deviation between both fields increases monotonically from $\Delta B\approx 3\%$ to $\Delta B\approx 49\%.$ (b)\,The curvature  $^\text{H}\!\kappa_\text{s}<^\text{f}\!\kappa_\text{s}$ within the annotated range $ h_\text{lob} < h < h_\text{upb}$.  The minimum with $\kappa_\text{s}\!=\!0$ is at $^{\text{H}}\!h_\text{opt}\!=\!1/\sqrt{6}\!=\!\,^{\text{H}}\!h_\text{st}$. The minimum of $^{\text{f}}\!\kappa_\text{s}$ is at $^{\text{f}}\!h_\text{opt}\approx0.576\neq \,^{\text{f}}\!h_\text{st}$.
}
\label{fig:8}
\end{figure}

To explore this in greater detail, we begin by examining the focused arrangement along $h\!=\!f$ (represented by the dashed yellow line in Fig.~\ref{fig:7}). Both the field and the curvature are plotted as a function of the distance $h$ from the ring in Fig.~\ref{fig:8}. These values can be compared to those of the Halbach configuration. The field in the focused arrangement is consistently stronger due to its design. Both, the relative and absolute difference between the two fields increase monotonically with increasing $h$, reaching a level of $49\%$ at $h\!=\!0.8$.

The curvatures $\kappa$ in Fig.~\ref{fig:8}(b) for $h\!=\!0$ are the same as the ones shown in Fig.~\ref{fig:4}(c). Increasing the distance $h$ leads to a first crossing of these two $\kappa_\text{s}$-lines at the lower boundary $h_\text{lob}\!\approx\!0.24$. A second crossing at higher values can be seen at the upper boundary $h_\text{upb}\!\approx\!0.49$. For a working distance $h$ between the lower boundary $h_\text{lob}$ and the upper boundary $h_\text{upb}$, the Halbach configuration has the better homogeneity, but its field $B_\text{x}$ is about 15\% smaller here than for the focused arrangement. Outside the range $h_\text{lob}\!<h\!<\!h_\text{upb}$, both the criteria, field strength and homogeneity, favor the height-adapted arrangement. 

The minimum of the curvature of the Halbach configuration is given by $\kappa_\text{x} = \kappa_\text{y} = \kappa_\text{s}\!=\!0$, which is reached at the distance of $^{\text{H}}\!h_\text{opt}\!=\!1/\sqrt{6}$, which coincides with the location of the steepest descent of $B_\text{x} (0,0,z)$ named $^{\text{H}}\!h_\text{st}$. 

The minimum of the curvature $\kappa_\text{s}$ of the focused configuration is located at $^{\text{f}}\!h_\text{opt}\!\approx\!0.576$. That value is very close to $1/\sqrt{3}$ , with a deviation in the \textperthousand  ~range. So, as a rule of thumb $^{\text{f}}\!h_\text{opt}\!\approx\!^{\text{H}}\!h_\text{st} \sqrt{2}$ can be used. It is remarkable that this optimal height is not located at the same place as the steepest descent of $B_\text{x} (0,0,z)$.  The curvature at $^{\text{f}}\!h_\text{opt}$ is finite, positive in the $y$ direction and negative in the $x$ direction, with $\kappa_\text{y} = -\kappa_\text{x} = \kappa_\text{s}\!\approx\!0.07$.

\begin{figure}[ht]
\includegraphics[width=.48\textwidth]{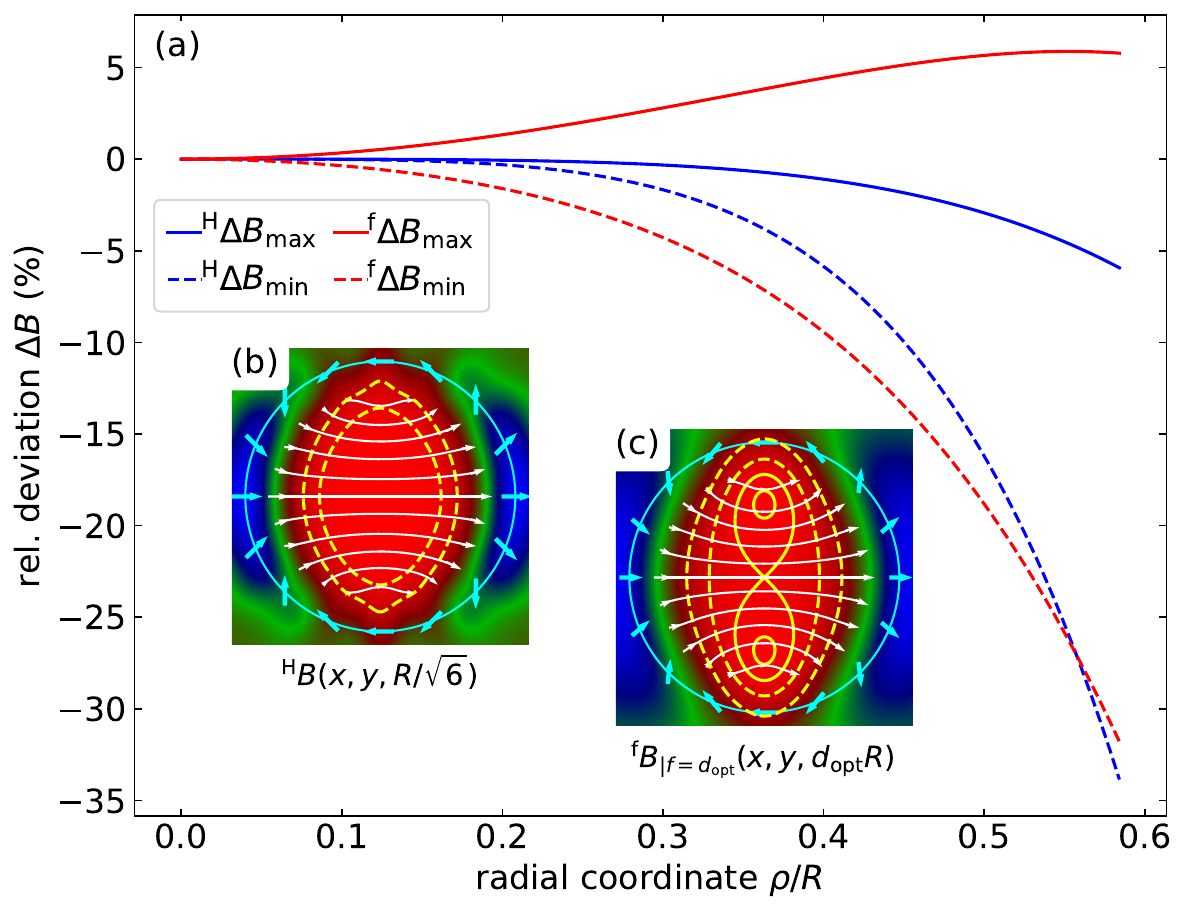}
\caption{Comparing field characteristics of the Halbach and the focused arrangement, with respectively minimized $\kappa_\text{s}$. (a)\:The deviation $\Delta B_\text{x}(\rho)$ changes with $4^\text{th}$  order of $\rho$ for the Halbach configuration, but only with $2^\text{nd}$ order for the focused one.  (b)\:The field of the Halbach configuration at $^{\text{H}}\!h_\text{opt} $, with the contour line for a -30\% and -10\% deviation from the center (dashed yellow). The cyan arrows show the projection of the dipoles located on a ring with $\rho\!=\!1$ (cyan circle).  (c)\:The field of the focused configuration for $f\!=\!h\!=\! ^{\text{f}}h_\text{opt}$. Contour lines for -40\% , -10\%(dashed), 0\%, and 5\% (solid yellow).} 
\label{fig:9}
\end{figure}

Figure~\ref{fig:9} presents an illustration of the two fields at two notable points highlighted in Fig.~\ref{fig:8}. The first field corresponds to the Halbach arrangement at a distance$^{\text{H}}\!h_\text{opt} $, where $\kappa_\text{s}$ is exactly zero. Both, the curves for the minimal and the one for the maximal deviation are of $4^{th}$ order near the center. However, the field is not rotationally invariant. The differences between the $x$ and $y$ directions are clearly visible for $\rho\!\gtrsim\!0.3$, and are also indicated by the shape of the -40\%-contour shown in the inset (b). 

The other field of interest is the one obtained with the focused arrangement at a distance $^{\text{f}}\!h_\text{opt}$.  It is shown in (c). It should be noted that the area where the field stays within a variation of 40\% is slightly larger compared to the one shown in (b). 
\subsection{\label{sec:theory_stacked}Stacked rings}
When the working distance from the dipole ring is finite, a sandwich configuration becomes a viable option. This denotes a configuration consisting of two dipole rings separated by a distance $d R$, with one ring positioned  below the working plane at $-d R/2$ and the other above that plane at $+d R/2$. If the upper dipole configuration is a mirror image of the lower one with respect to the working plane, no additional calculations are required to determine $B_x$, which is simply doubled due to the symmetric arrangement. Consequently, the values of $\kappa$ remain unchanged from those of the single ring. Therefore, Fig.~\ref{fig:8} continues to serve as a useful reference to guide the construction of such a sandwich of rings. This sandwich construction has the additional advantage that both $B_\text{z}$ and $\partial B_\text{z}/\partial z$ must be zero in the middle plane due to the mirror symmetry of the design.
\subsection{\label{sec:theory_twist}Twisted rings}
\begin{figure*}[t]
\onecolumngrid
\includegraphics[width=.99\textwidth]{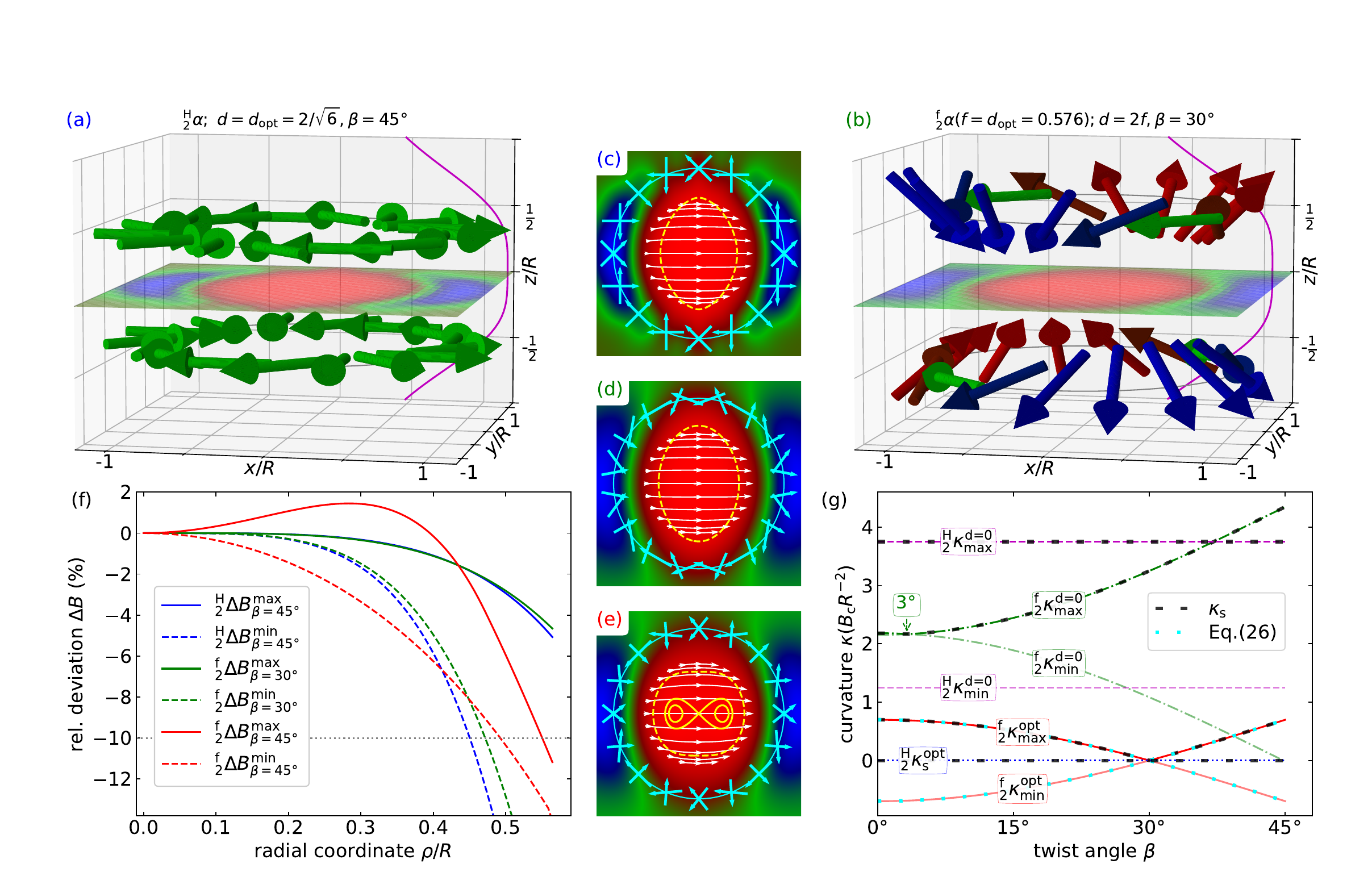}
\caption{The rings of Fig.~\ref{fig:9} in a twisted setup: (a) Two Halbach rings (green arrows) twisted by $\beta\!=\!45^\circ$. Their $B_\text{x}(0,0,z)$ field is shown in the background (magenta line, arb.\ units). The plane in the middle displays $B_\text{x}(x,y,0)$ (detailed view in (c)). (b) Two focused rings twisted by $\beta\!=\!30^\circ$. (c, d, e): $B_\text{x}(x,y,0)$, with  projections of the $2\times16$ dipoles (cyan arrows). Contour lines for -10\% (dashed), 0\%, and 1\% (solid yellow) field change. (c) is for the configuration shown in (a),  and (d) respectively for (b). (e): same as (d), but with $\beta\!=\!45^\circ$. (f) Field deviations $\Delta B_\text{x}(\rho)$ for the configurations (c, d, and e). (f) Influence of $\beta$ on $\kappa$ of 4 configurations (text boxes), with $\kappa_\text{s}$ (black short dashed) and Eq.~(\ref{eq:equal_kap}) (cyan dotted) as overlays.} 
\label{fig:11}
\twocolumngrid
\end{figure*}

With such a sandwich configuration, an intriguing possibility emerges. One ring can be rotated relative to the other around the $z$-axis, as shown in Fig.~\ref{fig:11}. If the upper ring is turned by $\beta$ in the counterclockwise direction and the lower ring (by -$\beta$) in the clockwise direction, this will lead to a reduction of $B_\text{x}$ by a factor of $\cos\beta$. The potential advantage of this approach is an improvement in field homogeneity, achieved by mitigating anisotropy, as illustrated in more detail in Fig.~\ref{fig:11}. Naturally, this adjustment also introduces a twist in the $B$-field vector along the $z$-axis. Whether this is an acceptable trade-off depends heavily on the specific application. For example, in magnetic resonance experiments, such a twist in the $B_{xy}$-field may not pose an issue, as the (radio-frequency) excitation field can still be aligned along the $z$-axis.

In Figure~\ref{fig:11}(a) two Halbach rings are twisted against each other with $\beta\!=\!45^\circ$, that is, the rings are oriented perpendicular to each other, as can be seen more clearly in Fig.~\ref{fig:11}(c). The distance $d\!=\!2/\sqrt{6}$ is selected as the optimal spacing to maximize homogeneity (cf. Fig.~\ref{fig:8}). Comparing the results in Fig.~\ref{fig:11}(c) with Fig.~\ref{fig:9}(a), the field of the Halbach configuration has not changed its quality due to the twist. The field has only increased by a factor of $\sqrt{2}$, compared to a single ring at a distance of $d/2$. Without twist, it would double, making it clear that twisting does not offer an advantage for this configuration at this specific distance. This can be explained by the fact that, in second order, the field near the center appears rotationally invariant, as indicated by the condition $\kappa_\text{min}\!=\!\kappa_\text{max}\!=\!0$. 

On a broader scale, the magnetic field in the Halbach configuration remains anisotropic, as highlighted in the motivation for this paper (cf. Fig.~\ref{fig:1}). Interestingly and counterintuitively, the introduction of a twist does not appear to significantly diminish this anisotropy. This observation also extends to the Halbach arrangement with $d\!=\!0$, as discussed in more detail in relation to Fig.~\ref{fig:11}(g) below.

In contrast, the focused arrangement undergoes significant changes when subjected to a twist. For Figs.~\ref{fig:11}(d, e) exactly the distance $d\!=\!2d_\text{opt}$ is chosen, where $\kappa_\text{min}\!=\! - \kappa_\text{max}$ (cf. Fig.~\ref{fig:8}).  

The magnetic field of such magnets at a twist angle of $\beta\!=\!30^\circ$ is shown in Fig.~\ref{fig:11}(d), which should be compared with Fig.~\ref{fig:9}(b).  For a quantitative comparison, the values of $\Delta B(\rho)$ from Fig.~\ref{fig:9}(c) and the ones from  Fig.~\ref{fig:11}(f) can be used. At this angle of twist, $\kappa_s$ equals zero, because the negative $\kappa_\text{min}$ perfectly cancels the positive $\kappa_\text{max}$ (cf. Fig.~\ref{fig:11}(g)). Consequently, the variation of the field strength from the center outward shows a dependence on the order $4^\text{th}$, as in the Halbach arrangement. 

A comparison in terms of $\kappa$ is not possible for these two special arrangements. Both $\Delta B_\text{x}(\rho)$- curves are of $4^\text{th}$ order here, and $\kappa_\text{min}\!=\!\kappa_\text{max}\!=\!\kappa_\text{s}=0$ applies for both configurations. For a meaningful comparison, an additional criterion is required. For example, considering the green and blue curves in Fig.~\ref{fig:11}(f), if a fixed deviation (e.g., 10\%) from the central field value is acceptable as such a criterion, the area where this condition is fulfilled would be approximately 10\% larger in the twisted focused configuration.

A twist angle of $\beta\!=\!45^\circ$ minimizes $xy$-anisotropy not only locally around the center but throughout the entire $B_\text{x}$ field. The result for the focused configuration is shown in Fig.~\ref{fig:11}(e). From a visual judgment, the effect appears to be the most appealing in this case.  The quantitative comparison in Fig.~\ref{fig:11}(f) reveals that the effect of this twist leaves $\kappa_\text{s}$ finite, but the 10\%-criterion clearly favors this configuration. It results in an approximately 20\% increase in the area of the so-defined "homogeneous" region compared to the optimal Halbach arrangement.

Figure~\ref{fig:11}(g) depicts the effect of the twist on anisotropy, represented by the curvatures for four configurations. Interestingly, for the Halbach configurations with distances $d\!=\!0$ and $d\!=\!2/\sqrt{6}$, both $\kappa_\text{min}$ and $\kappa_\text{max}$ remain completely unchanged. This surprising behavior is likely a result of the inherent symmetry of $^{\text{H}}\!\alpha$. 

For the focused configuration with a sandwich distance $d\!=\!0$ (that is, two rings in a plane), a tiny reduction of $\kappa_\text{s}$ can be achieved by a twist of approximately $3^\circ$. However, the effect is minimal and probably not worth the effort. 

For the sandwich with a distance $d\!=\!2f_\text{opt}\!\approx\!1.15$,  the curvature $\kappa_\text{min}\!=\!-\kappa_\text{max}$ for any angle of twist $\beta$.  We achieve $\kappa_\text{min}\!=\!\kappa_\text{max}\!=\!\kappa_\text{s}\!=\!0$ at a twist angle of $\beta\!=\!30^\circ$.

\begin{figure*}[t]
\onecolumngrid
\includegraphics[width=.99\textwidth]{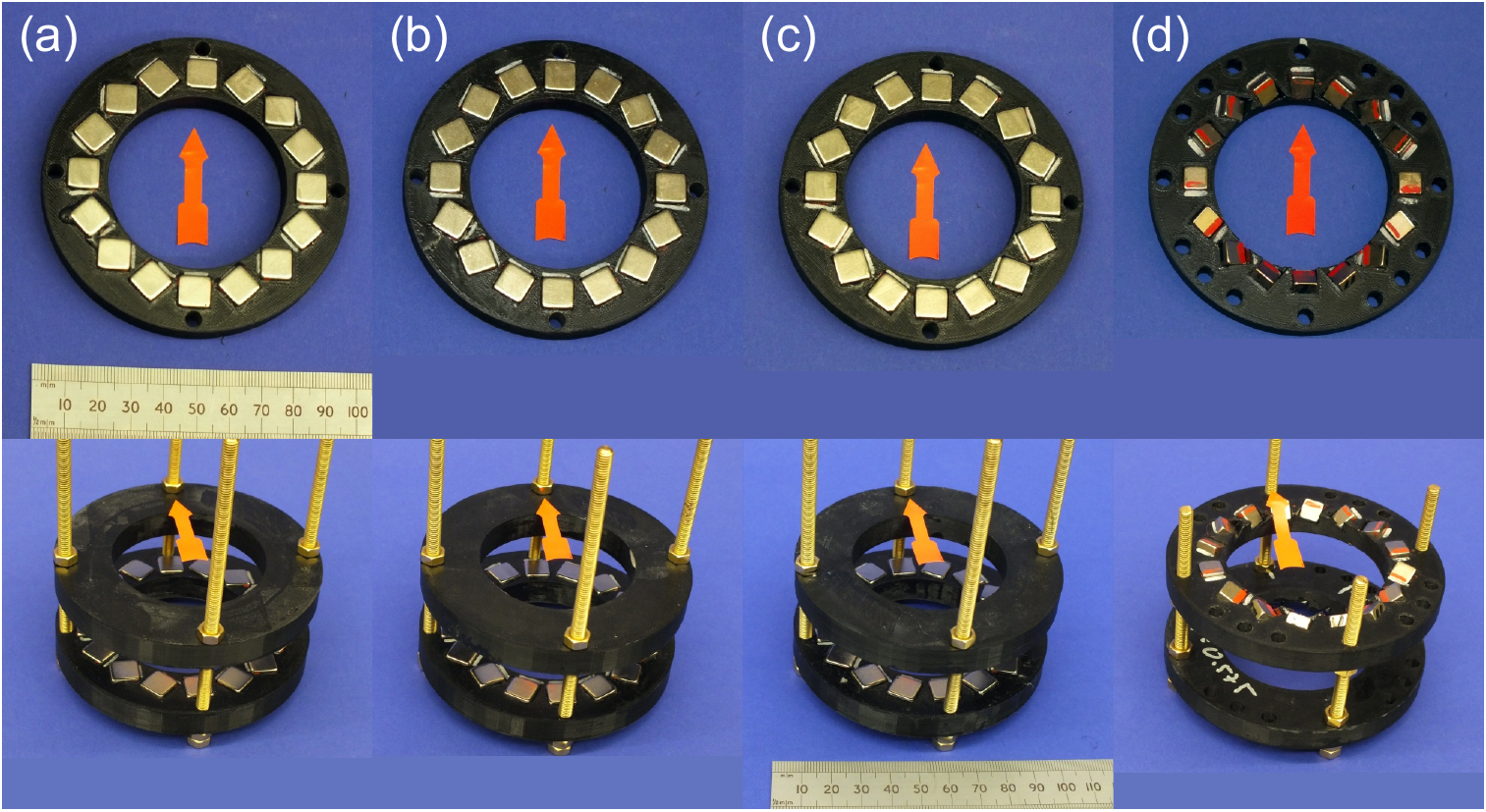}
\caption{Photographs of all constructed magnets. Single rings (top row) are stacked to a sandwich (bottom row). The red arrow indicates the field direction in the center. (a) Halbach configuration with $^\text{H}\alpha$ defined in Eq.~(\ref{eq:alpha_H}), (b) homogeneity optimized configuration with $^\text{h}\alpha$ defined in Eq.~(\ref{eq:8}), (c) field optimized configuration with $^\text{f}\alpha(0)$ defined in Eq.~(\ref{eq:alpha_B}), and (d) focused configuration $^\text{f}\alpha(-0.576)$ (i.\ e., focused below the surface) defined in Eqs.~(\ref{eq:alpha_f}) and (\ref{eq:theta_f}). In the stack, both rings are parallel (no twist angle).}
\label{fig:12}
\twocolumngrid
\end{figure*}
This can be understood by discussing the lowest order approximation for a field with the basic ingredient 
$\kappa_\text{x}\!=\!-\kappa_\text{y}$, which reads:
\begin{equation}\label{eq:model_field}
\qquad \qvec{B}(x_\text{r}, y_\text{r}) =  \begin{pmatrix}1-x_\text{r}^2/2+y_\text{r}^2/2\\x_\text{r}y_\text{r}\\ \end{pmatrix},
\end{equation}
where the index r refers to the coordinate system of the ring producing that field. The curvatures are thus $\kappa_\text{x$_\text{r}$}\!=\!-1$ and $\kappa_\text{y$_\text{r}$}\!=\!1$. The $y_\text{r}$-component is necessary to fulfill $\nabla \cdot \qvec{B}\,=\,0$. If one ring forming such a field is turned by $\beta$ and the other on by $-\beta$, the superposition of both fields can be written with the rotation matrix 
\begin{equation}
\qquad \qvec{R}_\beta =  \begin{pmatrix}\cos\beta & -\sin\beta\\\sin\beta & \cos\beta \end{pmatrix}
\end{equation}
as
\begin{equation}
\qquad \qvec{B}(x, y) = \qvec{R}_\beta^{-1}\cdot\qvec{B}\left(R_\beta\cdot \begin{pmatrix}x\\y\end{pmatrix}\right)+\qvec{R}_\beta\qvec{B}\cdot\left(R_\beta^{-1}\cdot \begin{pmatrix}x\\y\end{pmatrix}\right)
\end{equation}
yielding
\begin{equation}\label{eq:twist}
\qquad \qvec{B}(x, y) =  \begin{pmatrix}2\cos\beta -x^2\cos(3\beta)+y^2\cos(3\beta)\\2xy\cos(3\beta) \end{pmatrix}.
\end{equation}
Comparing (\ref{eq:twist}) with (\ref{eq:model_field}) we conclude
\begin{equation}\label{eq:equal_kap}
\kappa_\text{x}\!=\!-\frac{\cos(3\beta)}{\cos\beta}, \text{ and } \kappa_\text{y}\!=\!\frac{\cos(3\beta)}{\cos\beta}
\end{equation} 
for a stack of two rings with a twist angle $\beta$. Thus, the ratio of the two curvatures is independent of $\beta$, as illustrated in Fig.~\ref{fig:11}(g), where the appropriately scaled Eq.~(\ref{eq:equal_kap}) is plotted together with the twist-dependent curvatures obtained for the focused configuration. Moreover, both curvatures become zero for $\beta\!=\!30^\circ$.

\section{\label{sec:experiment}Experimental setup and results}
So far, we have discussed rings with $N$=16 magnets. For choosing $N$ for the experimental setup, it is important to clarify how the field and its homogeneity depend on the number of magnets. We refer to a graphical user interface (GUI) where these questions can be explored interactively for all configurations discussed in this paper~\cite{Rehberg2025}. This GUI illustrates that there is a strong $N$ dependence for line dipoles: The center forms a saddle point of order $N$, and deviations from the center field grow $\propto$ $\rho ^N$. This describes an extremely flat field near the center for $N$=16. In contrast, for point dipoles, the dependence on \( N \) is significantly weaker. They generally lead to a parabolic field profile, i.~e., finite curvature values $\kappa_\text{x}$, $\kappa_\text{y}$, and $\kappa_\text{z}$, provided that \( N \geq 2 \). As can be interactively visualized~\cite{Rehberg2025}, variations in these curvatures remain below the percent level for all even numbers with \( N \geq 10 \). The choice of $N$=16, the smallest power of two satisfying that constraint, is motivated more by practical considerations than by scientific necessity: it facilitates visual inspection of the Halbach ring, as angles such as $90^\circ$, $45^\circ$, and $22.5^\circ$ (i.~e., successive bisections of $90^\circ$) are easier to recognize. 

For the design, cubes were chosen over spheres, despite the latter's ability to produce an exact point dipole field, because they offer practical advantages in positioning and alignment. Differences in field geometry between cubes and spheres are very small at the distances relevant to our experiment and can be accurately accounted for~\cite{Camacho2013, Bjork2023, Sosa2024}. Consequently, all magnet types discussed in the previous section were experimentally fabricated by gluing 16 FeNdB cuboids onto supports that were 3D printed using polylactic acid (PLA), as shown in Fig.~\ref{fig:12}.
 
For Halbach-, homogeneity optimized- (cf.\ Eq.~(\ref{eq:8})), and the field optimized (cf.\ Eq.~(\ref{eq:alpha_B})) configurations, $10\!\times\!10\times\!10~\text{mm}^3$ cuboids (N45 from dodego GmbH, Waldlaubers\-heim, Germany) were used, with a remanence of $B_\text{R}\!=\!1.33 - 1.36~\text{T}$ as specified by the manufacturer.

The focused configuration (cf. Eq.~(\ref{eq:theta_f})) was realized on the same radius $R \!=\! 35\, \text{mm}$, but due to the necessary out-of-plane tilting, smaller magnets were required. These were fabricated from $8\!\times\!8\times\!8~\text{mm}^3$ cuboids (N50 from EarthMag GmbH, Dortmund, Germany) with $B_\text{R}\!=\!1.42 \pm 0.1~\text{T}$.

The magnetic moment of about 150 magnets of each size was determined by measuring their individual far field using a Hall probe (HMNA-1904-VF from LakeShore Cryotronics) and optimal arrangements were calculated by minimizing the equations given in the Appendix \ref{sec:appC}. Remanence $B_\text{R}$, magnetization $M$,  and magnetic moment $m$ are related by $B_{\text{R}}\!=\!\mu_0 M \!=\! \mu_0 m /V$ with $V$ as the volume of the magnet.

The magnetic field generated by these magnets is measured using another 3-axis Hall probe (MV2 from Metrolab, Plan-les-Ouates, Switzerland), which is positioned in the $x$, $y$, and $z$ directions by stepper motors.

\subsection{\label{sec:Plane_rings}Plane rings}

\begin{figure}[ht]
\includegraphics[width=.48\textwidth]{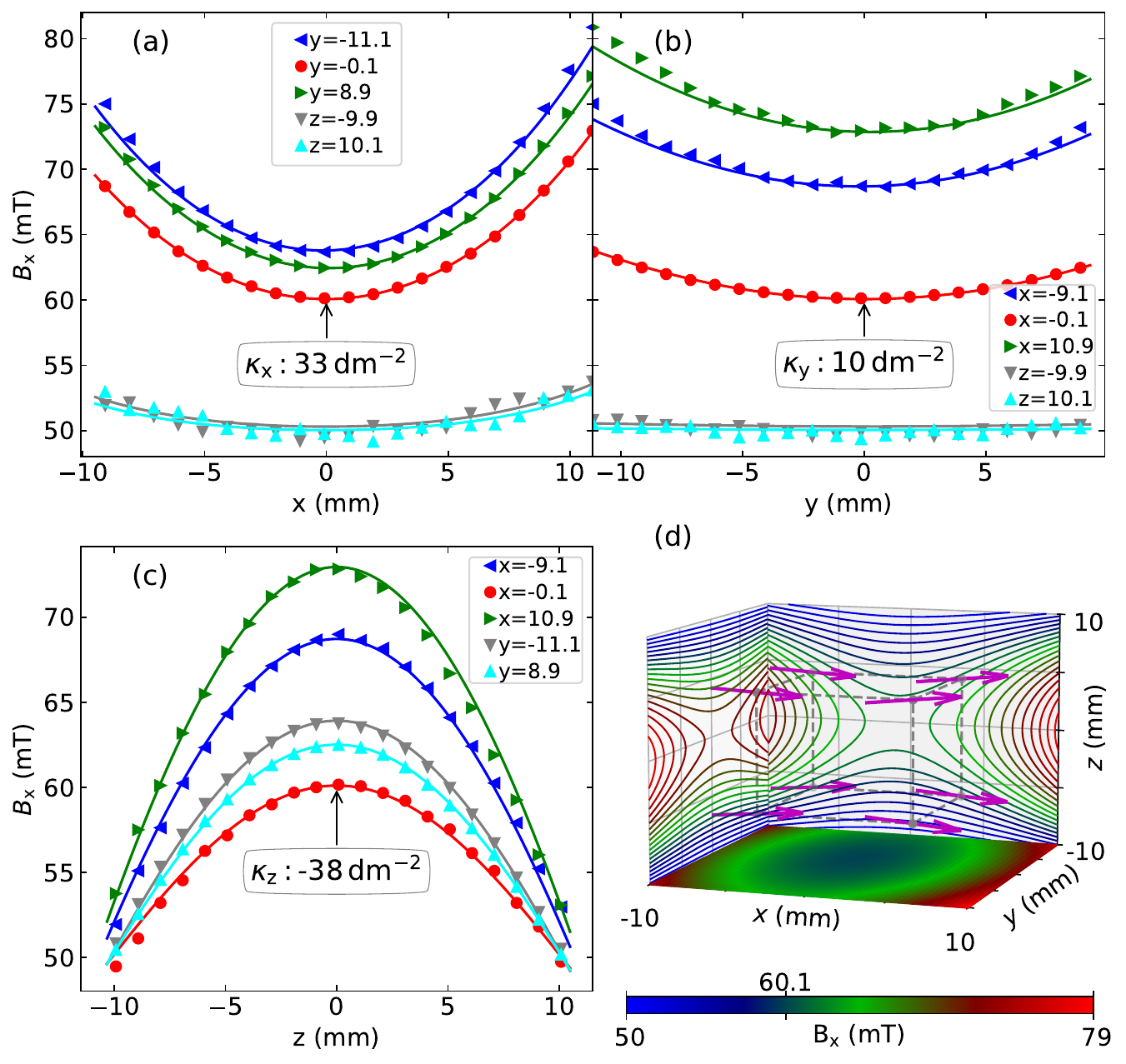}
\caption{Measurement of magnetic field in the Halbach configuration. A subset of the 9261 measurements within an 8 cm$^3$ cube is shown alongside the fitted values. (a) Measurement and fit along five lines parallel to the $x$-axes: the center line (red), 1 cm to the left (blue), to the right (green), below (gray) and above (cyan). The value of $\kappa_\text{x}$ is obtained from the fit to all 9261 data points. (b) Same as in (a), with lines parallel to the $y$-axis, and (c) parallel to the $z$-axis. (d) Eight measured $\qvec{B}$-vectors (magenta) positioned at the corners of a 1 cm$^3$ cube. At the bottom, $B_x$ in the $z\!=\!0$ plane is displayed as derived from the fit. The contour lines projected onto the walls correspond to the  $x\!=\!0$ and $y\!=\!0$ planes.} 
\label{fig:13}
\end{figure}

As a reference, the Halbach configuration measurement is shown in Fig.~\ref{fig:13}. It consists of 9261 data points collected within a $2\times 2\times 2~\text{cm}^3$ cube, with a step size of 1 mm. Five lines are shown in each direction: one passing through the center, along with the leftmost, rightmost, highest, and lowest lines. The coordinate system of the plots is centered with respect to the saddle point of $B_\text{x}$.  Its location is determined using the result of the fitting function, defined by the ansatz \begin{equation}\label{eq:fit}
B(x,y,z) = \sum_{i,j,k=0}^{i+j+k=2}a_{i,j,k}x^{2i} y^{2j}z^{2k},
\end{equation}
where the $a_{i,j,k}$ are the fit parameters. The shift from the coordinate system of the stepping motor by $x \!=\! x_\text{motor}-x_\text{c}$, etc., adds three additional ones. An ansatz with restriction to even powers of the coordinates is only justified when the coordinate system is parallel to $B_\text{c}$. This is accomplished by allowing for a small rotation of the coordinate system around the $z$-axis, thus adding another fit parameter.

Since the ansatz (\ref{eq:fit}) is based on a Taylor expansion around the origin, it is reasonable to assign a larger weight to data points closer to the center during the fitting process. To achieve this, the weights of the data points are reduced proportionally to their distance from the center.

\begin{figure}[ht]
\includegraphics[width=.48\textwidth]{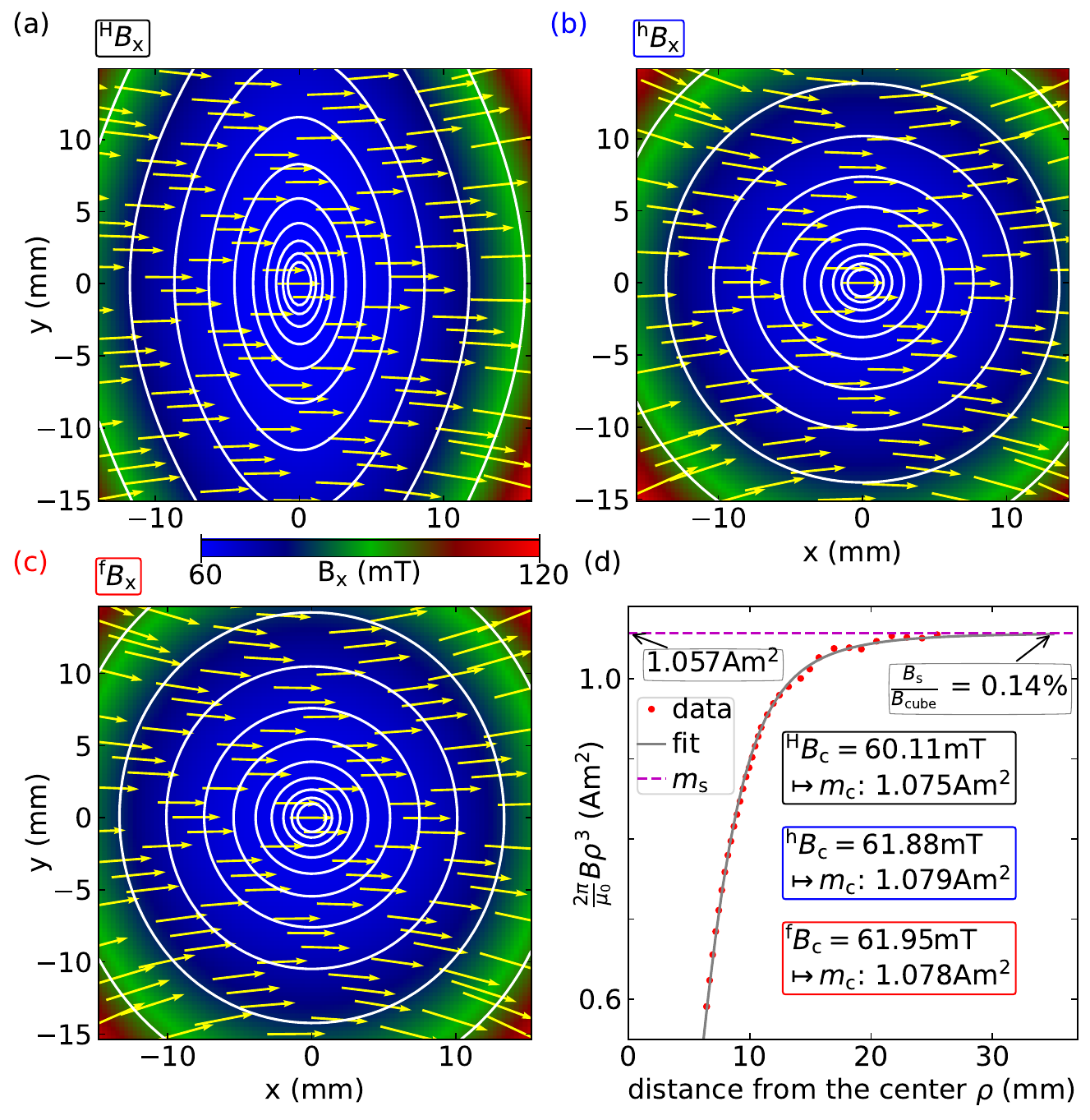}
\caption{$B_\text{x}$ measured in the $xy$-plane for three non-focused configurations: (a) Halbach configuration $\left(^{\text{H}}\!B_x(x,y)\right)$, with 1/8 of the measured 961 $(B_\text{x}, B_\text{y})$ data points plotted as yellow arrows. The colored background represents $B_\text{x}$ derived from the fit with white contour lines overlaid. The resulting field $B_\text{c}$ is listed in the text box in (d). (b) Same for the homogeneity optimized  configuration $\left(^{\text{h}}\!B_x(x,y)\right)$, and (c) for the field optimized configuration $\left(^{\text{f}}\!B_x(x,y)\right)$. (d) The determination of the magnetic moment of a cube, compared with the values derived from the fields in (a), (b), and (c). Only a quarter of the measured data are shown (red points). The asymptotic value of the curve yields the magnetic moment, while the magenta dashed horizontal line represents the corresponding value for a sphere.}
\label{fig:14}
\end{figure}

Figures~\ref{fig:13}(a-c) demonstrate that the fit accurately represents the field measurements near the center, allowing the extraction of the four key parameters:
$B_\text{c}=a_{0,0,0}=60.11 \pm 0.006~\text{mT}$, and the curvatures
$\kappa_\text{x}=2a_{2, 0, 0}/B_\text{c}$=$0.33\pm 0.003~\text{cm}^{-2}$,
$\kappa_\text{y}=2a_{0, 2, 0}/B_\text{c}=0.10\pm 0.003~\text{cm}^{-2}$, and
$\kappa_\text{z}=2a_{0, 0, 2}/B_\text{c}=-0.38~\pm 0.003~\text{cm}^{-2}$.
Their numerical values are also provided in Figs.~\ref{fig:13} and \ref{fig:14}. 
The lines for $z\!=\!-9.9~\text{mm}$ and $z\!=\!10.1~\text{mm}$ in (a, b) clearly illustrate that the field becomes flatter at greater distances from the ring, a topic explored in detail in \ref{sec:finite_dist}. 

Figures~\ref{fig:13}(a,b) are presented side by side without separation to emphasize that the data for each plane converge at a shared corner of the cube. Specifically, the uppermost point on the right-hand side of (a) is identical to the uppermost point on the left-hand side of (b), corresponding to $B_\text{x}(10.9 \text{ mm},\, -11.1 \text{ mm},\, 0.1 \text{ mm})$.  Similarly, the blue curve in (a) representing $B_\text{x}(x, -11.1 \text{ mm}, 0.1 \text{ mm})$, intersects with the green curve in (b), representing $B_\text{x}(10.9\,\text{mm},\, y,\, 0.1 \,\text{mm})$, as both belong to the same contour plane of the three-dimensional fit.

Figure~\ref{fig:13}(d) presents contour lines of the fitted function in the $x\!=\!0$ and $y\!=\!0$ planes, projected on the walls of the measuring cube. The fitted field in the $z\!=\!0$ plane is shown at the bottom in colors related to magnetic fields by the color bar below. This visualization highlights that the center field is a minimum within the $xy$-plane but a maximum along the $z$-direction, forming a saddle point. In addition, eight measured $B$-vectors at the vertices of a $1\text{ cm}^3$ cube are displayed as arrows. 

To facilitate a comprehensive comparison of the three different plane ("non-focused") configurations discussed in \ref{sec:comparison}, the measurement area in the $xy$-plane is expanded to $3\text{ cm}\times3\text{ cm}$. The results are shown in Fig.~\ref{fig:14}(a-c). The yellow arrows represent the measured fields within the plane, with only every eighth data point shown for the sake of clarity. The contour plots illustrate the fits to Eq.~(\ref{eq:fit}) in the $z=0$ plane. Ten contour lines are plotted, corresponding to geometrically increasing levels defined as $B_\text{x}/B_\text{c}-\!1 \!=\! 2^i$ \textperthousand, where $i$ denotes the number of the contour line counted from the center ($i\!=\!0$).

All quantitative aspects of the measurements presented here agree with the theoretical predictions in Fig.~\ref{fig:4}. The central field $B_\text{c}$ is lowest in the Halbach configuration and highest in the field-optimized configuration. Furthermore, the anisotropy observed in the Halbach configuration is significantly reduced in the modified versions.

The central fields $B_\text{c}$ are provided in the insets of Fig.~\ref{fig:14}(d). Using these values and the numerical results presented in Section~\ref{sec:comparison}, the magnetic moments $m$ of the sixteen point dipoles (or spheres) that make up the ring can be estimated under the assumption that they are all the same.  It is noted that the experimentally determined $m_\text{s}$ may differ slightly, as the rings were constructed from cubes rather than idealized spheres. 

The ratio of fields produced by cubes and spheres is analytically known as a function of distance $\rho$~\cite{Camacho2013, Sosa2024}. At a distance of $R\!=\!35\text{ mm}$, the field difference between a cube and a sphere is only 1.5\textperthousand, but even this tiny correction is taken into account when estimating the magnetic moments of the cube, $m_\text{c}$, as shown in the inset. The calculated moments have an average value of 1.077 Am$^2$ and deviate from this average by no more than 2\textperthousand. 
This value should be compared with an independent measurement of the magnetic moment $m_\text{c}$. This is done by measuring $B_\text{x}$ for a cube with its magnetic moment aligned along the $x$-direction. For visual inspection of the data, it is advantageous to multiply the measured $B_\text{x}$ by $\rho^3$, where $\rho$ is the distance from the center of the magnet. The ensuing curve should asymptotically approach a constant which, when multiplied by $2\pi/\mu_0$, gives the magnetic moment. For spheres, this procedure would give a straight horizontal line, as shown by the magenta dashed line in Fig.~\ref{fig:14}(d). This line also serves to highlight the asymptotic value of the measured curve.

The determination of the asymptotic value is applicable to magnets of any shape. However, since we are working with a cube, the corresponding theoretical curve can also be fitted, as shown by the gray solid line. As an additional note, this fit enables a precise determination of the cube's center relative to the position of the Hall probe, which is otherwise challenging to resolve at a sub-millimeter scale. Note that this is only true if uniform magnetization of the cube is assumed.

More importantly, the magnetic moment can be deduced as a fitting parameter. This method yields $m_\text{c}\!=\!1.057\pm 10^{-9}~\text{Am\(^2\)}$, where the error is estimated from the scatter of the residuals. Specifically, the standard deviation of the residuals is used to scale the covariance matrix returned by the fitting routine, as is done in all fits presented in this paper. This small statistical uncertainty is negligible compared to the dominant systematic errors, such as those arising from the determination of the cuboid volume and the calibration of the Hall probe. Considering that, the value is in good agreement with the value of 1.077 Am$^2$ obtained from the measurements presented in Fig.~\ref{fig:14}(a-c), within a 2\% deviation. The corresponding remanence of the $1\:\text{cm}^3$ cubes is $1.33\:\text{T}$, which is well within the technical specifications provided by the manufacturer.

\begin{figure}[ht]
\includegraphics[width=.48\textwidth]{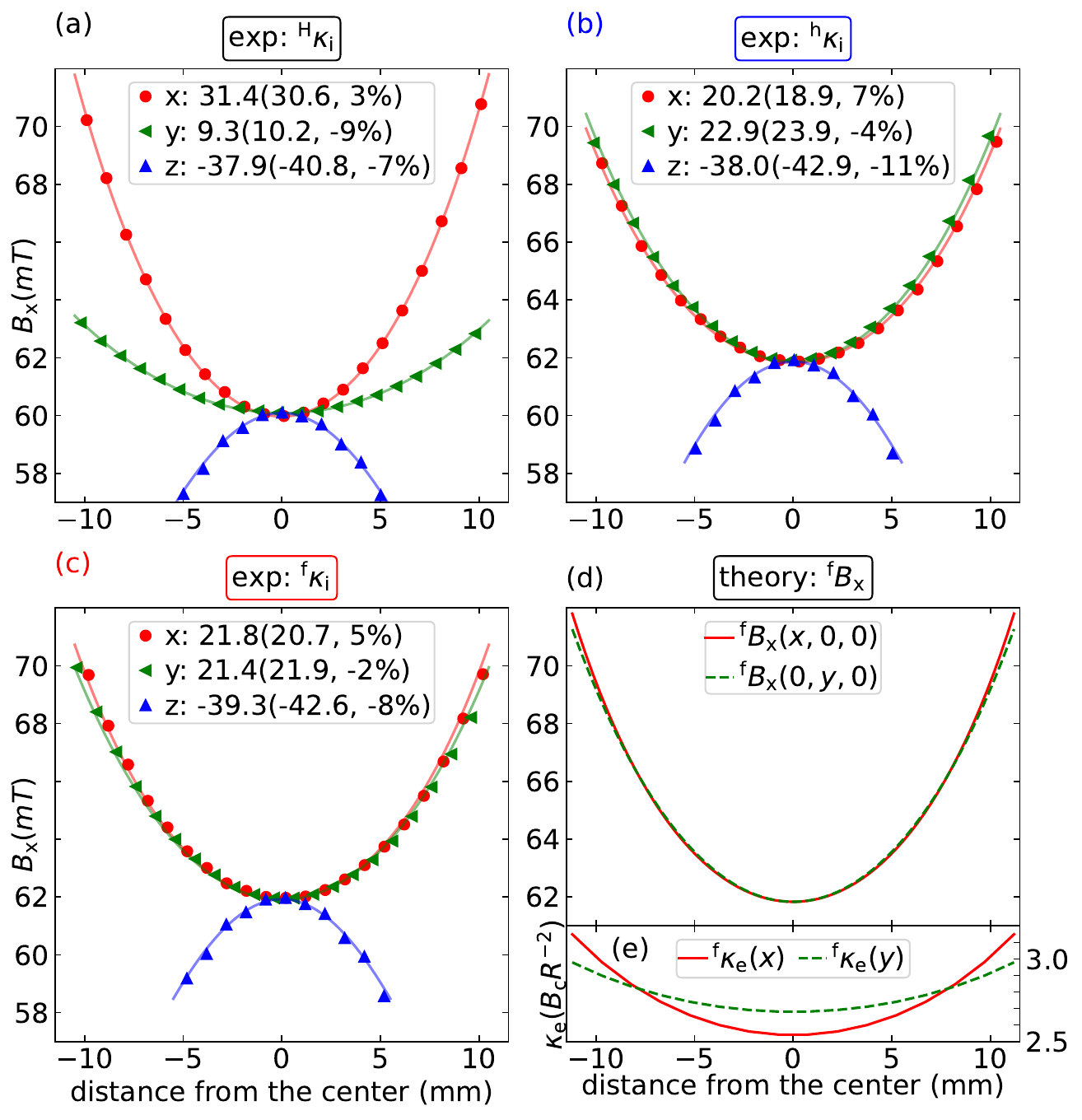}
\caption{The curvatures $\kappa$ obtained from 3D-measurements for three non- focused configurations: (a) Halbach ($^{\text{H}}\!\kappa_{x,y,z}$), (b) the homogeneity-optimized ($^{\text{h}}\!\kappa_{x,y,z}$) and (c) the field-optimized ($^{\text{f}}\!\kappa_{x,y,z}$) configuration. (d) Theoretical estimation of $^\text{f}B_\text{x}$. (e) The corresponding effective curvatures.}
\label{fig:15}
\end{figure}

In addition to the center field, the fit also provides the curvatures $\kappa$ in all directions. A quantitative comparison with the theoretically expected values for all three configurations is presented in Fig.~\ref{fig:15}.

The labels show the measured curvatures obtained from the fit of Eq.~\ref{eq:fit}. The numbers in brackets indicate the theoretically expected value, along with the deviation between the experimental and theoretical results. The agreement is within the range of 10\%.

However, there is one notable discrepancy for the field-optimized configuration: the measurements indicate a slightly larger curvature along the $x$-direction compared to the $y$-direction, whereas the theoretical expectation is the opposite. Although this effect is small, it appeared robust for different measurements (not shown here). The explanation is provided in Fig.~\ref{fig:15}(d), which shows the theoretically expected values for $^{\text{f}}\!B_x$ along the
$x$- and $y$--directions. The difference is tiny and thus difficult to detect with the resolution of this theoretical plot, not to mention the experimental one. However, it becomes clearly apparent in Fig.~\ref{fig:15}(e), which shows the effective curvature $\kappa_\text{e}$ as introduced in Fig.~\ref{fig:4}. For radii smaller than $7\,\text{mm}$, the curvature along the $x$-direction is expected to be smaller than along the $y$-direction. However, at larger distances from the center, $^\text{f}\kappa_\text{e}(x,0,0)$exceeds $^\text{f}\kappa_\text{e}(0,y,0)$, which is then in agreement with the measurements shown in Fig.~\ref{fig:15}(c). 

\subsection{\label{sec:focused_configuration}Focused configuration}
\begin{figure}[ht]
\includegraphics[width=.48\textwidth]{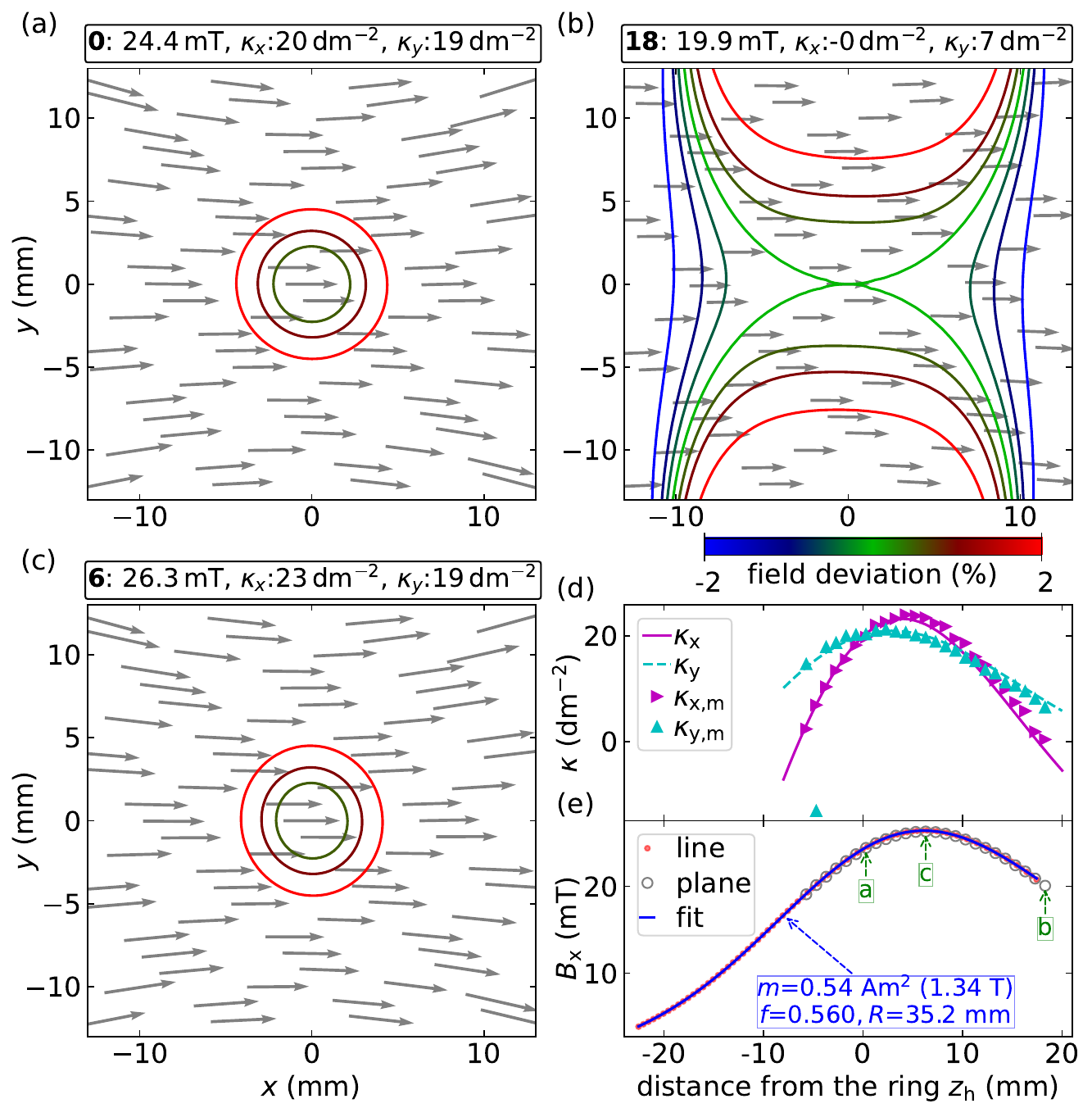}
\caption{The magnetic field of the focused configuration. (a-c) $B_\text{x}$ and $B_\text{y}$ (gray arrows) as measured in the planes with $z_\text{h}$=0, 6 and 18 mm. The contour lines indicate levels of equal deviation in steps of 0.5\% from the center value. Their color is set according to the color bar. (d) The curvatures measured in the planes (triangles), and the theoretical expectation (lines). (e) $B_\text{x}(0,0,z)$ as measured on a line along $z$ (red dots), and a second measurement (gray circles) obtained from planes as shown in (a-c), and the fitted theory (solid blue curve). The numbers in the text box list the fit parameters.}
\label{fig:16}
\end{figure}

For the focused configuration,  smaller magnets $8\!\times\!8\times\!8~\text{mm}^3$ had to be used on the ring of $R\!=\!35\ \text{mm}$ to allow for a sufficiently large tilting angle of the cuboid magnets. The focus length was chosen as $f\!=\!0.576$, which corresponds to a distance of 20.1 mm. Fig.~\ref{fig:16} shows the resulting magnetic field in 3 different planes, one within the ring, the other 6 mm and 17 mm apart from the ring. The gray arrows represent the $x$- and $y$-components of the measured magnetic field. The most striking feature is the fact that the largest center field, namely $B_\text{c}\!=\!26.0 \ \text{mT}$, is not obtained within the plane of the ring, but rather at a distance of 6 mm. This distance must be compared with the numerically obtained expectation provided in Eq.~(\ref{eq:h_max}),
which yields 6.4 mm. 

For distances $z_\text{h}\!=\!0\ \text{mm}$ and $z_\text{h}\!=\!6\ \text{mm}$ the field $B_\text{c}\!=\!B_\text{x}(0,0,z_\text{h})$ is a local minimum.  Three contour lines are plotted around this center, corresponding to an increase of 0.5\%, 1\% and 2\% above this center field $B_\text{c}$.

At a distance of $z_\text{h}\!=\!18\ \text{mm}$,  $B_\text{x}$ forms a saddle point at the center, as shown in Fig.~\ref{fig:16} (b). The contour line where $B_\text{x}\!=\!B_\text{c}$ is shown in green. The red line indicates the value $B_\text{x}\!=\!1.02\,B_\text{c}$, as indicated by the color bar. It is obvious that the field in this plane is the most homogeneous of the three planes shown here: within the area between the red and blue lines, the deviation is less than 2\%. That area is much larger than the ones within the red ellipses of Figs.~\ref{fig:15} (a) and (c). 

A quantitative comparison with the theoretical expectation for this focused arrangement is presented in Figs.~\ref{fig:16} (d) and (e). In the lower part, the measured field $B_\text{x}(0,0,z_\text{h})$ is fitted by the theoretical curve for this arrangement, where the magnetic moment $m$, the radius of the ring $R$, and the focus length $f$ are the fit parameters. The magnetic moment $m\!=\!0.54\,\text{Am}^2$ corresponds to a remanence of 1.34 T for $8\!\times\!8\times\!8~\text{mm}^3$ cubes, which is within the range $B_\text{R}\!=\!1.42 \pm 0.1~\text{T}$ provided by the manufacturer. 

The value obtained for $R$, 35.2 mm, with a deviation of 0.2 mm between the fitted value and the one expected from the design, is well within the precision of the setup.

The deviation in the focus length $f$, 0.560,  is less than 3\% from the desired value $f=0.576$.  This deviation roughly corresponds to a deviation in the tilt angle $\theta$ of less than 2 degrees according to Eq.~(\ref{eq:theta_f}). 

The measured values for the curvatures $\kappa_\text{x}$ and $\kappa_\text{y}$ are presented in Fig.~\ref{fig:16} (d). The solid lines representing the theory are not fitted but were calculated with the values for $f$ and $R$ listed in Fig.~\ref{fig:16} (e). Considering that fact, the agreement is convincing. 

\subsection{\label{sec:stacks}Stacked rings}

\begin{figure}[ht]
\includegraphics[width=.48\textwidth]{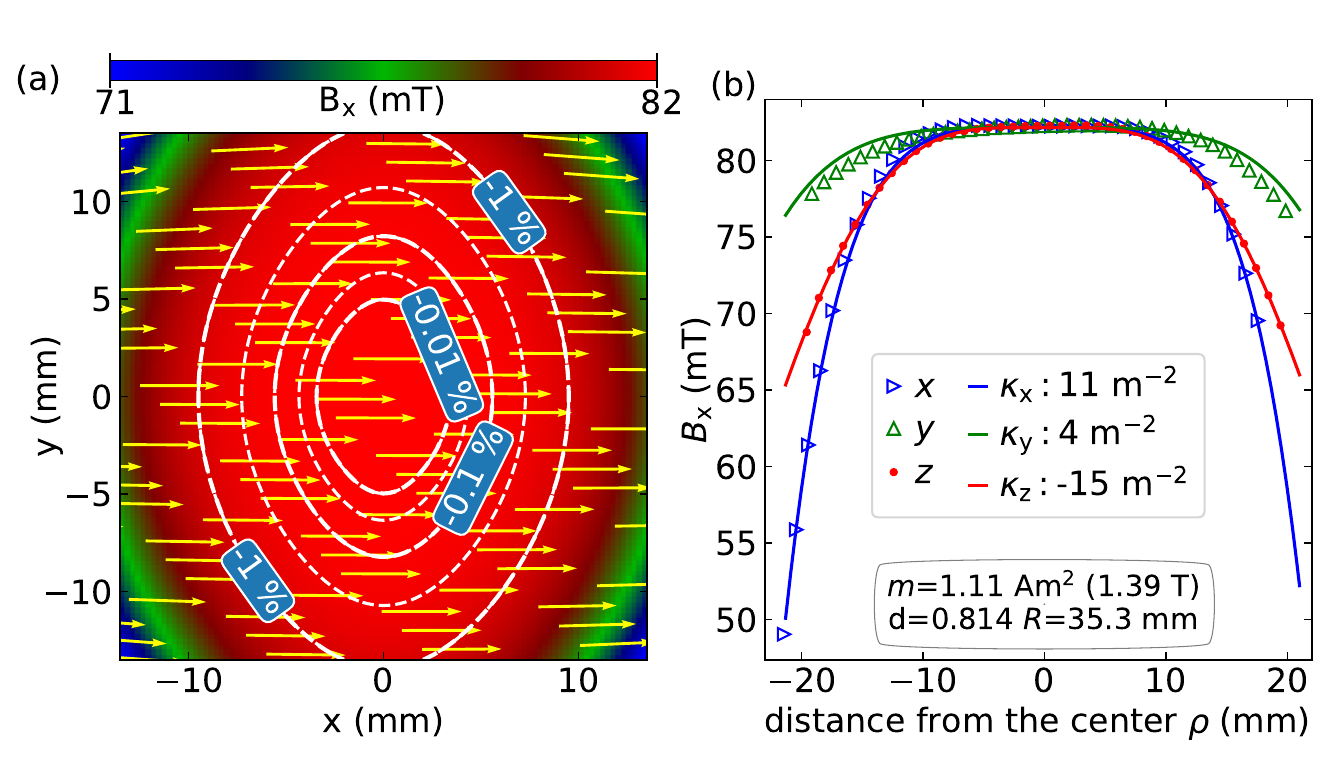}
\caption{Field measurements within a Halbach stack. (a)~The field in the  middle plane (yellow arrows). The scale of $B_\text{x}$ is given by the background color (color bar). White contour lines are shown for up to 1\% deviation. (b) $B_\text{x}$ measured along the $x, y,$ and $z$-coordinate. The solid lines denote a common fit yielding the parameters listed in the boxes.}
\label{fig:17}
\end{figure}

As discussed in the theoretical part, the homogeneity of the field in a plane can be increased by making use of stacks. This is demonstrated for the Halbach configuration close to the optimal distance $d_\text{opt}$ in Fig.~\ref{fig:17} (distances always refer to the centers of the magnets in the two rings). 

It is obvious that the absolute values of $\kappa$ become small here, they fall below 1/dm$^2$. Thus, the measured curves in Fig.~\ref{fig:17}(b) appear to be of $4^\text{th}$ order by visual inspection. The fact that $\kappa_\text{x}$ and $\kappa_\text{y}$ are slightly positive here is barely visible because the negative $4^\text{th}$ order terms take over at a distance of about 3 mm from the center. As a consequence of the small values of $\kappa_\text{x}$ and $\kappa_\text{y}$, the range where the field deviation is smaller than 1\% is approximately 4 cm$^2$ as indicated by the corresponding contour line in Fig.~\ref{fig:17}(a). These lines still appear elliptical rather than round, similar to those shown in Fig.~\ref{fig:1}(b). This is in agreement with expectations because the ratio $\kappa_\text{x}/\kappa_\text{y}\! \approx \!3$ for any distance from the Halbach ring according to Fig.~\ref{fig:8}. However, it must be stressed that $\kappa_\text{s}$ is reduced by a factor of about 50 in this stack configuration, when comparing it with the measurement shown in Fig.~\ref{fig:14}(a) and Fig.~\ref{fig:15}(a).

\begin{figure}[ht]
\includegraphics[width=.48\textwidth]{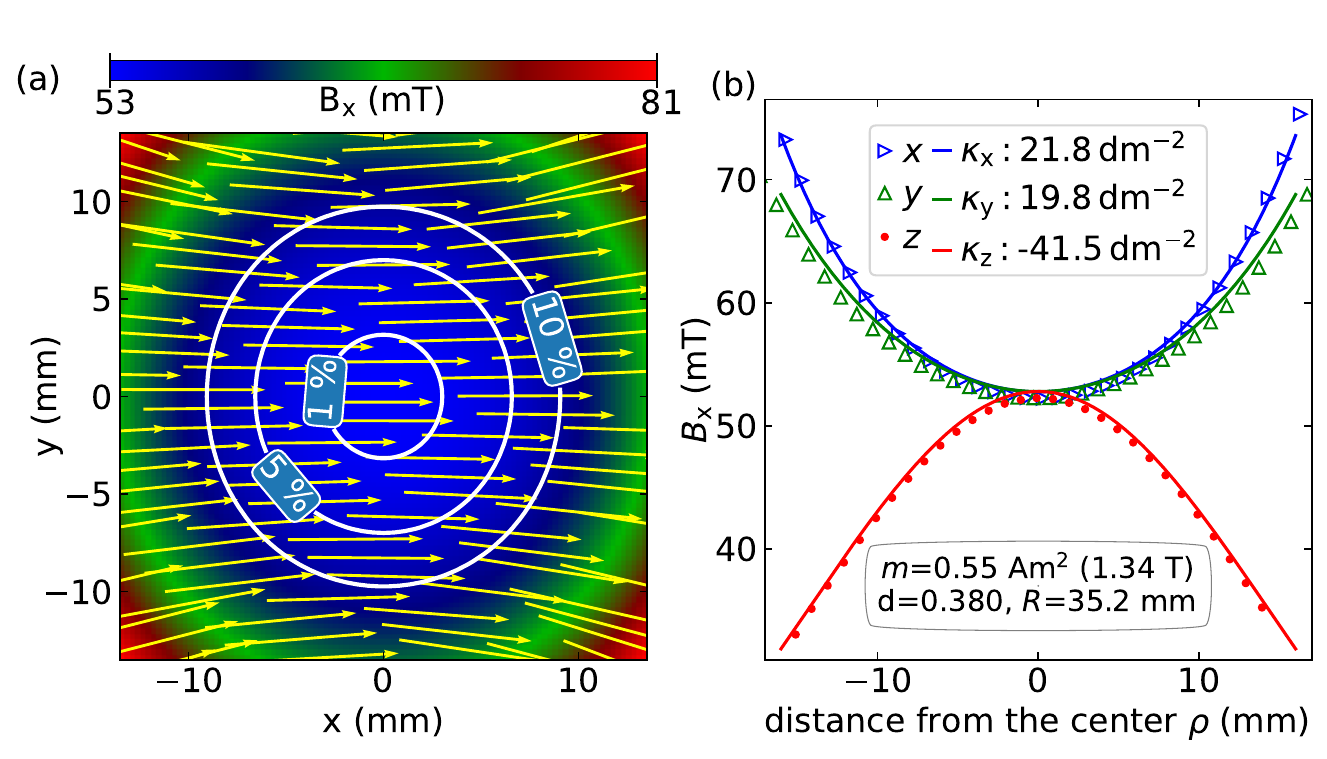}
\caption{Two rings of the focused configuration stacked at a distance of 12.5 mm. (a) The field measured in the plane where $B_\text{c}$ has a maximum (yellow arrows.) The background color represents the strength of $B_\text{x}$. Contour lines indicate the deviation from $B_\text{c}$. (b) The field measured along the $x,y,z$-direction. The curves are fits of the theoretical configuration with $d$ as a common fit parameter. The other parameters (text box) are taken from the measurements in Fig.~\ref{fig:16}.}
\label{fig:18}
\end{figure}

The disadvantage of anisotropic curvature is reduced with the focused configuration, and this is also true when building stacks from such rings, as shown in the measurements presented in Fig.~\ref{fig:18}. Here, two focused rings were stacked, as shown in Fig.~\ref{fig:12}(d). The distance between the rings was adjusted to 12.5 mm (although a submillimeter precision is hard to achieve here). At this distance, the field along the $z$-axis should be close to a maximum. With the values for $f\!=\!0.576$ and $R\!=\!35~\text{mm}$ taken from the design of the ring, Eq.~(\ref{eq:h_max}) yields a distance of 12.4~mm. 

Inspecting Fig.~\ref{fig:18} shows an almost isotropic minimum of $B_\text{x}$ in the $xy$ plane and a clear maximum along the $z$ axis. When we compare the absolute values of the center field with the one shown in Fig.~\ref{fig:17}, it might be striking that the values are lower here. However, this is due to the fact that the magnets in these rings are smaller, namely 8 mm instead of 10 mm. Thus, we expect a reduction by the factor $(8/10)^3\!\approx\! 0.5$ here. With the same size and number of magnets the field would indeed be larger with the focused stack at this distance.

\begin{figure}[ht]
\includegraphics[width=.48\textwidth]{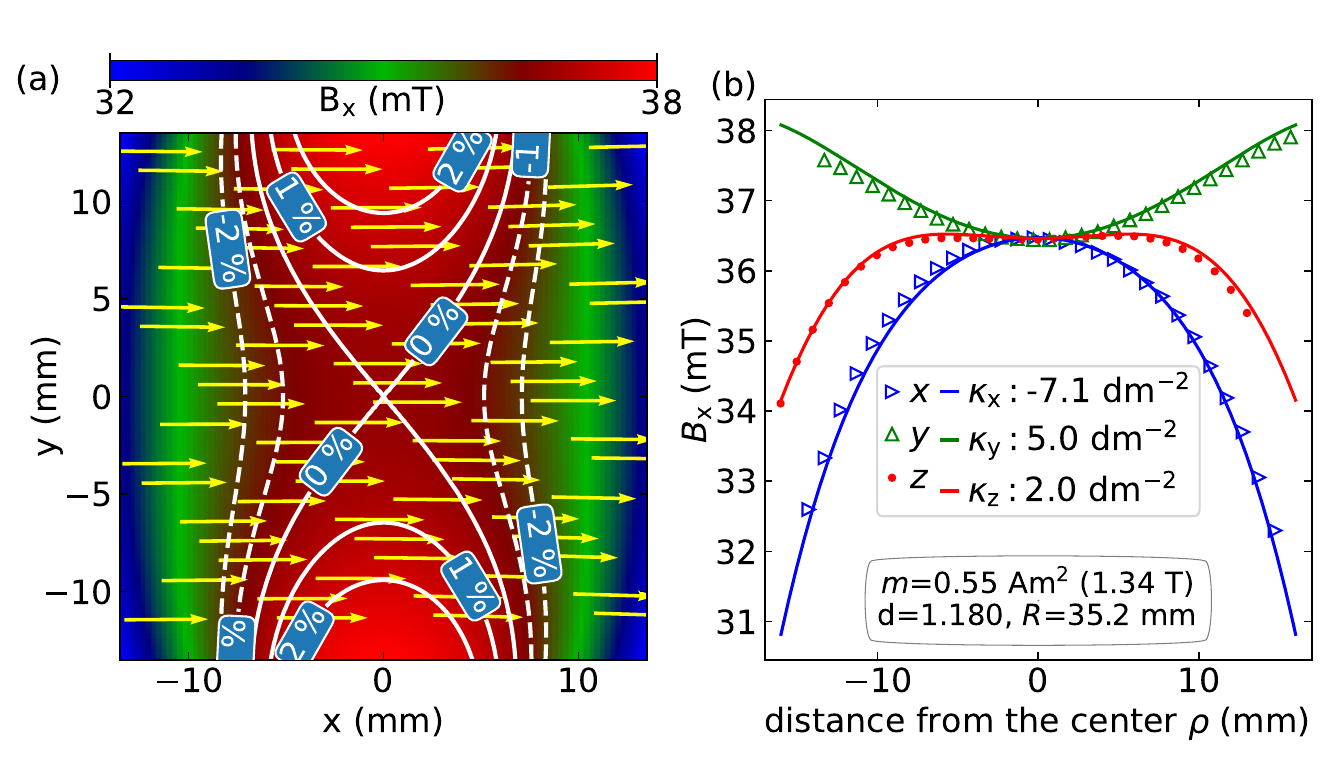}
\caption{Two rings of the focused configuration stacked at a distance of 41 mm. (a) The field measured in the center plane (yellow arrows.) The background color represents the strength of $B_\text{x}$. Contour lines indicate the deviation from $B_\text{c}$. (b) The field measured along the $x,y,z$-direction. The curves are fits of the theoretical configuration with $d$ as a common fit parameter. The other parameters are the same as in Fig.~\ref{fig:18}.}
\label{fig:19}
\end{figure}

A measurement of two focused rings stacked with a distance close to the optimal one,  $d_\text{opt}$,  is shown in Fig.~\ref{fig:19}.  Note that the relation $\kappa_\text{min}\!=\!-\kappa_\text{max}$ serves to define this distance, as illustrated in Fig.~\ref{fig:8}.
The measurement was taken on a stack with a distance between the centers of the magnet rings of 41 mm. The fit to the profiles yields $d=1.176$,  which is slightly larger than $2d_\text{opt}=1.152$. As a consequence, the absolute value of the positive $\kappa_\text{x}$ must be higher than the absolute value of the negative $\kappa_\text{y}$ (cf. Fig.~\ref{fig:8}), which is the case for the measurements shown here.

Working at such a saddle point has a practical advantage, as illustrated in Fig.~\ref{fig:18}(b). Because the lines of zero deviation from the center value $B_\text{c}$ extend throughout the plane, one obtains a substantial increase in the homogeneous area with a variation within the limits of $\pm 2\%$.
\subsection{\label{sec:twisted stacks}Twisted stacks}
\begin{figure}[ht]
\includegraphics[width=.48\textwidth]{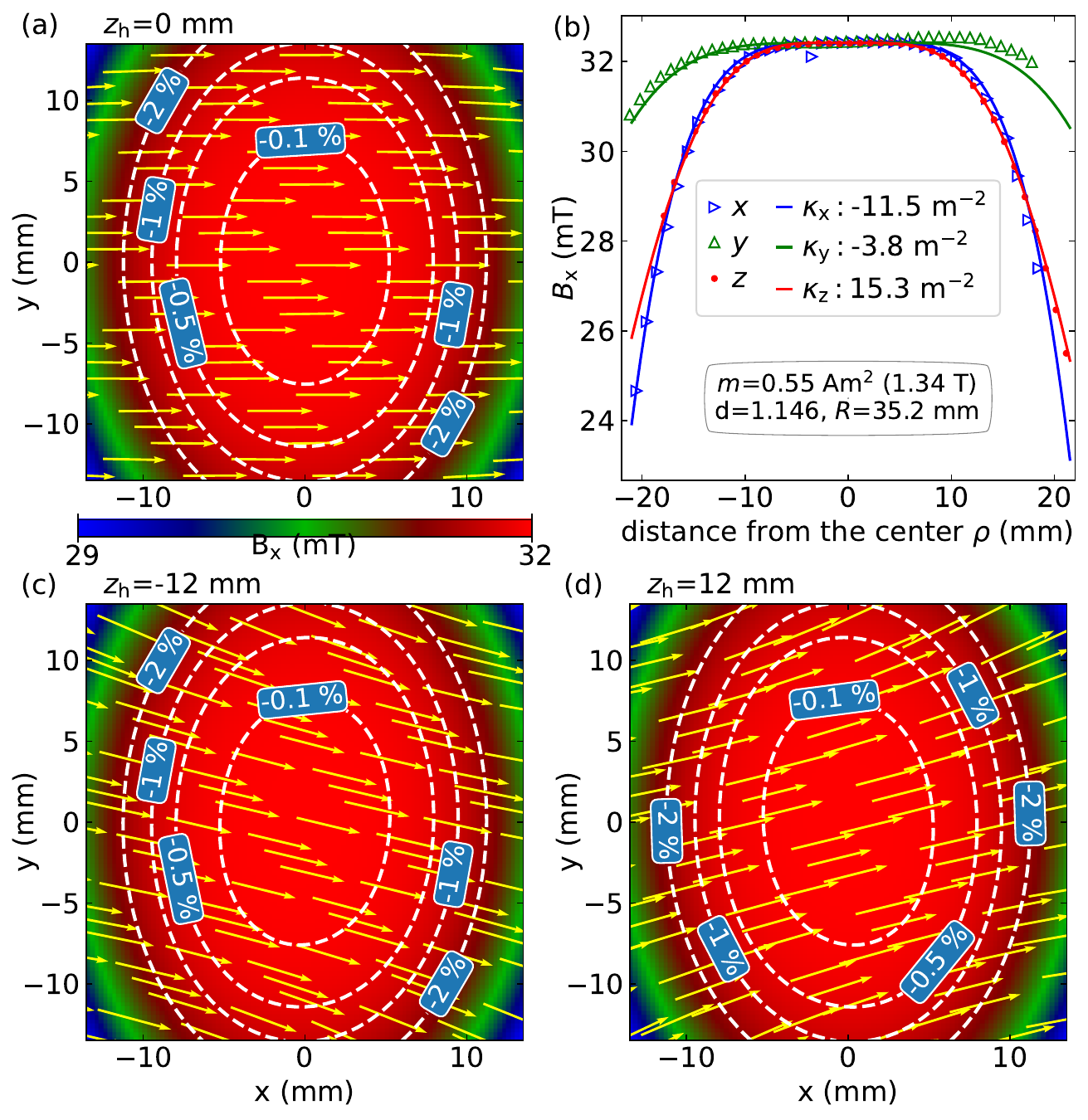}
\caption{Same as Fig.~\ref{fig:19}, but twisted by $\beta=30^\circ$.  (a) Field in the middle plane.  (b) Same as Fig.~\ref{fig:19}. One data point from the $x$-profile seems distorted. (c) Field measured in a plane 12 mm below (d) and above the center.}
\label{fig:20}
\end{figure}

Following theoretical considerations, it should be possible to increase the homogeneous area by introducing a twist in the stacked configuration where the two curvatures are of opposite sign. This applies particularly for the distance $d=2h_\text{opt}$ and a twist angle of $\beta=30^\circ$. An experimental setup has been built to match these conditions illustrated in Fig.~\ref{fig:11}(b), and the result is shown in Fig.~\ref{fig:20}.

The curvatures are below $1\,\text{dm}^{-2}$ and as such so tiny that they cannot be directly seen in Fig.~\ref{fig:20}(b), as the homogeneity is now governed by fourth-order terms. As a result, the range where the deviation from the center field is smaller than 1\% is about $3\,\text{cm}^2$. 

To illustrate the twist of the magnetic field, Figs.~\ref{fig:20}(c,d) display measurements in planes 12 mm below and above the center.  As the lower ring is turned clockwise around the $z$-axis, and the upper one counterclockwise, so is the magnetic field.

\section{Conclusion and Outlook}
Novel arrangements for the generation of homogeneous fields in a plane are presented, both as an analytical concept and as an experimental realization. These innovative design concepts address limitations of the classical Halbach configuration, particularly when constructed from short magnets, which deviate from the original design principles. In particular, the experimental results demonstrate a curvature reduction of more than 2 orders of magnitude. 

Analytical models are presented for the new designs that involve turns, tilts, and twists, which enhance both field strength and homogeneity. These considerations are restricted to dipoles arranged in a circle with a fixed radius. Natural extensions of this setup could involve deforming the ring within the plane or shifting the magnets into the third dimension~\cite{Tewari2023}. The experimental demonstration presented here employs a relatively coarse segmentation of the analytical (continuous) magnetization using cubic magnets. Remarkably, the discrete nature of the ring setup, consisting of only 16 cuboids, does not appear to influence the field. This can be attributed to the rapid decay of the modulation toward the ring's center. However, more advanced segmentation schemes~\cite{Insinga2016, Insinga2019} could be implemented to further enhance field homogeneity.

The focused design is especially interesting because it shifts the maximum field out of the magnet plane, making it highly suitable for single-sided magnetic resonance applications. With such a setup, the idea of a twist to minimize the curvature of the field could be used without actually involving a twist in the field. This could be achieved by two interwoven rings, one turned clockwise and the other in counterclockwise direction against the axis of the field. Furthermore, this approach enables the construction of magnet stacks with greatly improved performance in terms of field strength and uniformity.

Some of these configurations can be explored interactively using open-source animation software~\cite{Rehberg2024, Rehberg2025}. The software also demonstrates that the Halbach arrangement of point dipoles has no net dipole moment and that its field decays with the fifth power of the distance. In contrast, the other configurations discussed here possess a net dipole moment, and their fields consequently decay with the third power of the distance. Although this difference does not significantly impact the homogeneity considered here, it may be relevant for certain technical applications, particularly those involving the relative rotation of two or more of such rings.
\begin{acknowledgments}
Its a pleasure to thank Eric Aderhold, Günter Auernhammer, Thomas Friedrich, and Reinhard Richter for stimulating discussions. The authors appreciate the partial support for their collaboration from TRR146 (DFG grant no. 233630050).
\end{acknowledgments}

\appendix
\section{\label{sec:appA} Point- and line-dipoles}
\begin{figure}[ht]
 \includegraphics[width=.48\textwidth]{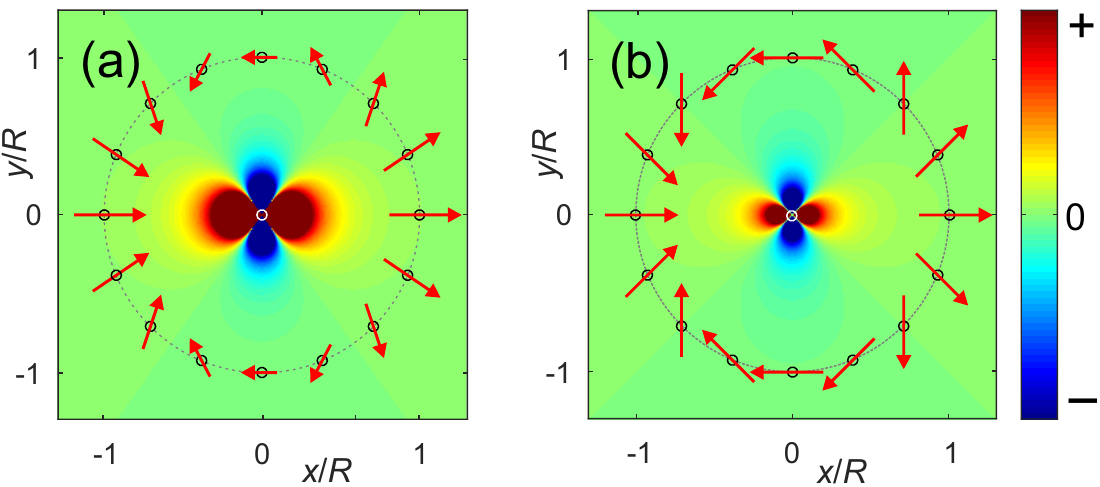}
\caption{Point vs. line dipole: (a) The colored image shows $B_x^{\text{point}}$ of a point dipole located at the origin (white circle). The red arrows show the direction of its magnetic field in a circle of radius $R$ around it. The length of the arrows depicts the magnitude of the local field. All in arbitrary units. (b)~Same as (a),  but for a line dipole.}
\label{fig:A1}
\end{figure}

The magnetic field is given by Eq.~(\ref{eq:B_dip}) for a point and by Eq.~(\ref{eq:2}) for a line dipole. For a location in polar coordinates $(R,\varphi,0)$, with  $\qvec{m}$ aligned along $x$ and located at the origin, this yields~\cite{Coey2002, Coey_book}:
\begin{equation*}
\qvec{B}^\text{point} = \frac{\mu_0 m}{4\pi R^3} \begin{pmatrix} 2\cos\varphi \\ \sin\varphi\\ 0\\ \end{pmatrix} , \quad  \qvec{B}^\text{line} = \frac{\mu_0 m'}{4\pi R^2} \begin{pmatrix} \cos\varphi \\ \sin\varphi\\ 0\\ \end{pmatrix}.
\end{equation*}
This difference between point and line dipoles is illustrated in Fig.~\ref{fig:A1}. In particular, the magnetic field along a circle around a line dipole is uniform in strength, as all arrows in Fig.~\ref{fig:A1}(b) have the same length.

\section{\label{sec:appB_new} A note on Eq.~(\ref{eq:alpha_B})}
To understand the choice of the correct $c$ for a smooth representation of the solution of Eq.~(\ref{eq:alpha_B}), it is helpful to note that the denominator becomes zero at the roots of the 2\textsuperscript{nd} Legendre polynomial, the so-called \textit{magic angle} $\varphi_\text{m} \!=\! \cos^{-1}(1/\sqrt{3})$. 
A smooth and unique term for the maximum angle $^{\text{f}}\!\alpha(0)$ is provided by choosing (cf.~Fig.~\ref{fig:2}a):
\begin{equation}\label{eq:12}
\begin{array}{l l l}
   {c \,=\, 0} & {\text{for}} & {\qquad\quad 0 \leq \varphi < \varphi_{\text{m}}}\\
   {c \,=\, 1} & {\text{for}} & {\qquad\,\, \varphi_{\text{m}} < \varphi <  \pi - \varphi_{\text{m}}}\\
   {c \,=\, 2} & {\text{for}} & {\,\,\, \pi-\varphi_{\text{m}} < \varphi <  \pi + \varphi_{\text{m}}}\\
   {c \,=\, 3} & {\text{for}}  & {\,\,\, \pi+\varphi_{\text{m}} < \varphi <  2\pi - \varphi_{\text{m}}}\\
   {c \,=\, 4} & {\text{for}} & {2\pi-\varphi_{\text{m}} < \varphi \leq  2\pi.}\\
\end{array}
\end{equation}

\section{\label{sec:appC} Optimal arrangement of individual Halbach and focused rings}
Since the permanent magnets used for the assembly of such rings have deviations in their magnetic properties (typically about 1~\%) it is advantageous to determine the individual deviation $({\Delta}m_{xj},{\Delta}m_{yj},{\Delta}m_{zj})$ and arrange the magnets so that a set of sums is minimized. This optimization procedure is described in detail in Appendix E in~\cite{Soltner2023} for Halbach multipoles. However, since the equations are presented here in tabular form, the expressions for dipoles (i.~e., $^\text{H}\alpha$) are repeated in Table \ref{tab:C1} for clarity.\\

\begin{table} [h]
    \setlength{\tabcolsep}{10pt}
    \renewcommand{\arraystretch}{1.5}
    \caption{List of eight sums whose absolute values need to be minimized to optimize magnet arrangements as a classical Halbach dipole (\ref{eq:alpha_H}).
    Hence, $\min\left\{\left| \sum_{j=1}^{N}{\text{expression}_j(m)}\right| \right\}$ for $m = $ I ...VIII must be fulfilled, compare to Eq.~(35) in~\cite{Soltner2023}.}
    \begin{flushleft}
    \begin{tabular}{ll}
        $m$&$\text{expression}_j(m)$ \\
        \hline
        I&$\Delta m_{zj}$ \\ 
        II&$\sin\varphi_j\;{\Delta}m_{zj}$\\ 
        III&$\cos\varphi_j\;{\Delta}m_{zj}$ \\ 
        IV&$\cos\varphi_j\;{\Delta}m_{xj}-\sin\varphi_j\;{\Delta}m_{yj}$ \\
        V&$\sin\varphi_j\;{\Delta}m_{xj}+\cos\varphi_j\;{\Delta}m_{yj} $ \\
        VI&$\left[\cos\varphi_j+3\right]\;{\Delta}m_{xj}-\sin\varphi_j\;{\Delta}m_{yj}$ \\
        VII&$\sin\varphi_j{\Delta}m_{xj} +\left[3\cos\varphi_j-3\right]{\Delta}m_{yj}$ \\
        VIII&$\cos\frac{\varphi_j}{2}{\Delta}m_{xj}-\sin\frac{\varphi_j}{2}{\Delta}m_{yj}$ \\
    \end{tabular}
    \label{tab:C1}
    \end{flushleft}
\end{table}

Totally analogously (procedure and nomenclature), Table \ref{tab:C2} gives eight equations to optimize the arrangement for magnets with orientation angles ${^\text{h}}\!\alpha$.

\begin{table}[h]
    \setlength{\tabcolsep}{10pt}
    \renewcommand{\arraystretch}{1.5}
    \caption{Optimization scheme for the $^\text{h}\alpha$-configuration (\ref{eq:8}). \\
     Substitutions: $\beta_j \equiv \tfrac{\pi}{8}\sin(2\varphi_j)$ and $\xi_j \equiv \beta_j-2\varphi_j$.}
    \begin{flushleft}
    \begin{tabular}{ll}
        $m$&$\text{expression}_j(m)$ \\
        \hline
        I&$\Delta m_{zj}$ \\
        II&$\sin\varphi_j\;{\Delta}m_{zj}$\\
        III&$\cos\varphi_j\;{\Delta}m_{zj}$ \\
        IV&$\cos\xi_j\;{\Delta}m_{xj}+\sin\xi_j\;{\Delta}m_{yj}$ \\
        V&$\sin\xi_j\;{\Delta}m_{xj}-\cos\xi_j\;{\Delta}m_{yj} $ \\
        VI&$\cos\xi_j\;{\Delta}m_{xj}+\sin\xi_j\;{\Delta}m_{yj}$ \\
        VII&$\left[3\sin\beta_j-\sin\xi_j\right]{\Delta}m_{xj}-\left[ 3\cos\beta_j-\cos\xi_j \right]{\Delta}m_{yj}$ \\
        VIII&$\left[3\cos\beta_j+\cos\xi_j\right]{\Delta}m_{xj}+\left[ 3\sin\beta_j+\sin\xi_j \right]{\Delta}m_{yj}$ \\
    \end{tabular}
    \label{tab:C2}
    \end{flushleft}
\end{table}

Table \ref{tab:C3} then refers to magnets with orientation angles $^{\text{f}}\!\alpha(f\!=\!0)$ (\ref{eq:alpha_B}), and Table \ref{tab:C4} gives the equations for focused magnets with ${^\text{f}}\!\alpha$ and $^\text{f}\theta$ as defined by Eq.~(\ref{eq:alpha_f}) and (\ref{eq:theta_f}).

\begin{table}
    \setlength{\tabcolsep}{10pt}
    \renewcommand{\arraystretch}{1.7}
    \caption{Optimization scheme for the $^\text{f}\alpha (0)$-configuration (\ref{eq:alpha_B}).\\
    Substitution: $\Xi_j \equiv \sqrt{5+3\cos\varphi_j}$.}
    \begin{flushleft}
   \begin{tabular}{ll}
        $m$&$\text{expression}_j(m)$\\
        \hline
        I&$\Delta m_{zj}$ \\
        II&$\sin\varphi_j\;{\Delta}m_{zj}$\\
        III&$\cos\varphi_j\;{\Delta}m_{zj}$ \\
        IV&$\Xi_j\;{\Delta}m_{xj}$ \\
        V&$\frac{1}{\Xi_j}\left(\vphantom{\frac{}{}}2\cos\varphi_j{\Delta}m_{xj}-\sin\varphi_j{\Delta}m_{yj}\right)$ \\
        VI&$\frac{1}{\Xi_j}\left(\vphantom{\frac{}{}}3\sin(2\varphi_j){\Delta}m_{xj}-4\;{\Delta}m_{yj}\right)$ \\
        VII&$\left[\vphantom{\frac{}{}}1+3\cos(2\varphi_j)\right]{\Delta}m_{xj}-3\sin(2\varphi_j){\Delta}m_{yj}$ \\
        VIII&$\left[\vphantom{\frac{}{}}1+3\cos(2\varphi_j)\right]{\Delta}m_{yj}+3\sin(2\varphi_j){\Delta}m_{xj}$ \\
    \end{tabular}
    \label{tab:C3}
    \end{flushleft}
\end{table}
\begin{table}[ht]
    \setlength{\tabcolsep}{10pt}
    \renewcommand{\arraystretch}{1.7}
    \caption{Optimization scheme for the focused configuration with ${^\text{f}}\!\alpha (f)$ from Eq.~(\ref{eq:alpha_f}) and  $^\text{f}\theta (f)$ from Eq.~(\ref{eq:theta_f}).\\
    Substitutions: \\ 
    $\Theta_j \equiv \cos^\text{f}\theta_j\;{\Delta}m_{xj} +\sin^\text{f}\theta_j\;{\Delta}m_{zj}$,\\
    $\widetilde\Theta_j \equiv \sin^\text{f}\theta_j\;{\Delta}m_{xj} -\cos^\text{f}\theta_j\;{\Delta}m_{zj}$,\\
    $\Phi_j \equiv 3\cos(^{\text{f}}\!\alpha_j-2\varphi_j)+\cos(^{\text{f}}\!\alpha_j)$, \\
    $\widetilde\Phi_j \equiv 3\sin(^{\text{f}}\!\alpha_j-2\varphi_j)+\sin(^{\text{f}}\!\alpha_j)$.}
    \begin{flushleft}
    \begin{tabular}{ll}
        $m$&$\text{expression}_j(m)$\\
        \hline
        I&$\widetilde\Theta_j $ \\
        II&$\sin\varphi_j\;\widetilde\Theta_j$\\
        III&$\cos\varphi_j\;\widetilde\Theta_j$ \\
        IV&$\cos(^{\text{f}}\alpha_j)\Theta_j - \sin(^{\text{f}}\alpha_j){\Delta}m_{yj}$ \\
        V&$\sin(^{\text{f}}\alpha_j)\Theta_j + \cos(^{\text{f}}\alpha_j){\Delta}m_{yj}$ \\
        VI&$\cos^\text{f}\theta_j\ \Phi_j\;{\Delta}m_{xj} - \widetilde\Phi_j\;{\Delta}m_{yj} +\sin^\text{f}\theta_j\Phi_j\;{\Delta}m_{zj} $ \\
        VII&$\widetilde\Phi_j\Theta_j+\left[\Phi_j-\cos(^{\text{f}}\alpha_j) \right]\;{\Delta}m_{yj}$ \\
        VIII&$\cos(^{\text{f}}\alpha_j-\varphi_j)\Theta_j - \sin(^{\text{f}}\alpha_j-\varphi_j)\;{\Delta}m_{yj}$ \\
    \end{tabular}
    \label{tab:C4}
    \end{flushleft}
\end{table}
\section{\label{sec:appC_new} Considerations on the focus length for a given working distance}
When designing an experimental setup, external conditions may determine the working distance from the dipole ring. This raises the question of how to select the focus to achieve optimal results. There is no unique answer, as Fig.~\ref{fig:7} made clear that maximal strength and minimal curvature of the field cannot be achieved simultaneously. 

\begin{figure}[h]
\includegraphics[width=.48\textwidth]{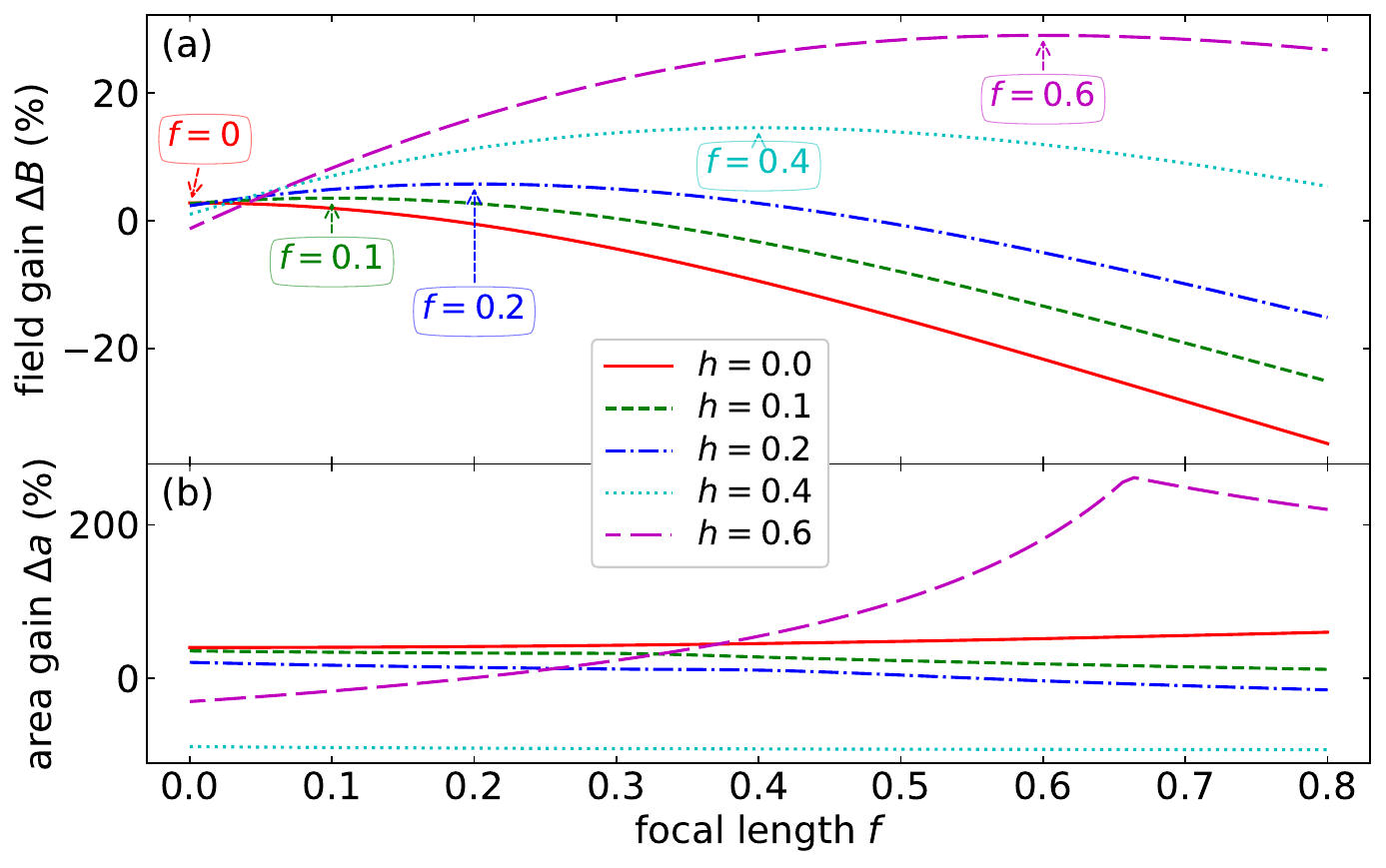}
\caption{
The influence of the focal length $f$ on the field and its homogeneity for five different heights $h$ (legend). (a) The field gain $ \Delta B$. Arrows point to the maxima located at $f\!=\!h$.  (b)~The area gain $\Delta a$ as defined in Fig.~\ref{fig:4}.} 
\label{fig:10}
\end{figure}

Figure~\ref{fig:10} is intended to provide a guide here. For five working distances $h$ it shows both the strength of the field and a measure of its homogeneity, as a function of the focal length. The field is quantified by its gain $ \Delta B\!=\!^{\text{f}}\!B_\text{x}(0,0,h) /^{\text{H}}\!B_\text{x}(0,0,h) -1$ with respect to the Halbach configuration. The homogeneity is quantified as the area gain $\Delta a\!=\!^{\text{H}}\!\kappa_\text{s}(0,0,h) / ^{\text{f}}\!\kappa_\text{s}(0,0,h) -1$ (cf.~Fig.~\ref{fig:4}).

Working at $h\!=\!0$, i.e., in the plane of the dipole ring, reveals for $f\!=0\!$ two quantities already seen in Fig.~\ref{fig:4}: The field gain is about 3\%, and the area gain is about 40\%. Interestingly, increasing the focal length to finite values further enlarges this area. However, this comes at the cost of reduced field strength.

For the working distances of $h\!=\!0.1$ and $h\!=\!0.2$, the field reaches its maximum at $f\!=\!h$. However, increasing $f$ results in a monotonic decrease in the area gain $\Delta a$ at these two distances.

The working distance $h\!=\!0.4$ falls into the range between $h_\text{lob}$ and $h_\text{upb}$ (cf.\ Fig.~\ref{fig:8}) where the Halbach arrangement has greater homogeneity. This is evident from the value of $\Delta a$ of approximately -100\% shown in Fig.~\ref{fig:10}(b), which is scarcely compensated for by the increase 10\% of the field $B_\text{x}$.

The situation is qualitatively different for the distance $h\!=\!0.6$. Here, the field increases by about 30\%, accompanied by a $\approx200\%$ enlargement of the homogeneous area. The plot of $\Delta a$ indicates that setting the focus to $f\!=\!0.66$, slightly above the working distance $h\!=\!0.6$, would be advantageous. The reduction in the field would be only 1\%, as the $\Delta B$-curve is quite flat near its maximum. Since lifting the homogeneous field area outside the region occupied by the magnets is highly relevant for single-sided magnetic resonance applications~\cite{Casanova2011, Casanova2016, Hobson2024}, a set of practical equations for the estimation of the associated magnetic fields is summarized in the Appendix~\ref{sec:appB}.

\section{\label{sec:appB} Central field of a focused arrangement of point dipoles}
\begin{figure}[ht]
\includegraphics[width=.48\textwidth]{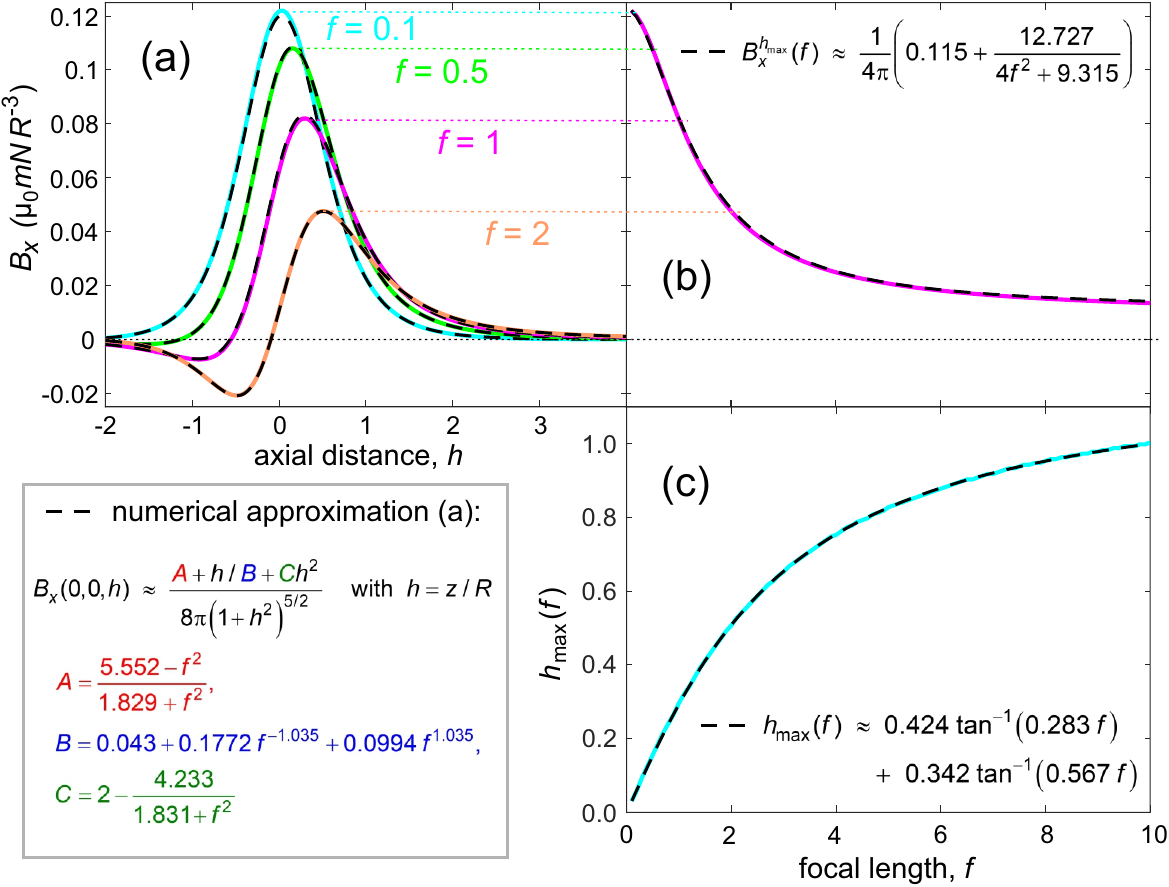}
\caption{(a) $B_x$-component of the magnetic field along the central axis, $z$, of a focused ring in units of its radius, $h = z/R$. The colored lines are calculated by Eq.~(\ref{eq:B1}) for $N\!=\!100$, the black dashed lines are the numerical approximations of Eq.~(\ref{eq:B2}). This was done for four values of $f$ (cyan: $f\!=\!0.1$, green: $f\!=\!0.5$, magenta: $f\!=\!1$, and orange: $f\!=\!2$). (b) The $B_x$-field at $z_{\text{max}}$ and comparison with Eq.~(\ref{eq:B4}). The four values of $f$ are indicated by dotted lines in the according colors. Both graphs, (a) and (b), share the same ordinate.
(c) The position of the axial field maximum $h_{\text{max}}(f) = z_{\text{max}}(f)/R$ versus $f$ and comparison with Eq.~(\ref{eq:B3}). Both graphs, (b) and (c), share the same abscissa. The composition of the numerically approximated curves (black dashed) is summarized in the lower left inset for a) and inserted into b) and c).}
\label{fig:B1}
\end{figure}
The focused arrangement of point dipoles shifts the maximal central field outside of the plane of the dipoles, which is particularly interesting for single-sided NMR~\cite{Casanova2011, Casanova2016}. Therefore, a few properties of such focused rings are discussed in this appendix.

$N$ dipoles on a circle with radius $R$ in the focused arrangement produce a field component $B_x$ along the $z$ axis. So, Eq.~(\ref{eq:dip_mom}) can be rewritten and rearranged by the powers of $h$ as
\begin{align}\label{eq:B1}
   \nonumber
   B_x& (0,0,hR) = \frac{\mu_0 m}{8\pi R^3(1+h^2)^{\frac{5}{2}}} \times \\
   \nonumber
   &\times \sum_{j=1}^{N}\left[{\cos^\text{f}\!\theta_j\left(3\cos(^{\text{f}}\!\alpha_j-2\varphi_j) +\cos^{\text{f}}\!\alpha_j\right) +} \right.\\
   &\qquad\quad+ 6h\cos\varphi_j\sin^\text{f}\!\theta_j -
   \left.2h^2\cos^{\text{f}}\!\alpha_j\cos^\text{f}\!\theta_j  \right].
\end{align} 

An analytical expression for the integration of $\varphi$ from 0 to $2\pi$ could not be obtained, so a complete dipole ring could only be numerically evaluated by the three coefficients to $h$ on their dependence on $f$. The following fit function was guessed:
\begin{align}\label{eq:B2}
  \nonumber
   B_x& (0,0,hR) \,\approx\, \frac{\mu_0 m N}{8\pi R^3(1+h^2)^{\frac{5}{2}}}
   \left[ \frac{a_1-f^2}{a_2+ f^2} + \right. \\
  & +\left. \frac{h}{b_1+b_2 f^{-b_3}+b_4 f^{b_3}} +
  \left( 2-\frac{c_1}{c_2+f^2}\right) h^2 \right],
\end{align} 
with the fitted coefficients $a_1\!=\!5.552$, $a_2\!=\!1.829$, $b_1\!=\!0.0427$, $b_2\!=\!0.17715$, $b_3\!=\!1.0346$, $b_4~\!=\!0.09943$, $c_1\!=\!4.233$, and $c_2\!=\!1.831$ obtained for $N$=100.

In the range $0<f<10$, this approximation has an average error of about 0.8 $\%$. Figure~\ref{fig:B1}(a) shows the excellent agreement between Eq.~(\ref{eq:B1}) with the approximation in Eq.~(\ref{eq:B2}) for four values of $f$. The field is asymmetric with a maximum at positive $z$ values.
Figure~\ref{fig:B1}(b) shows the dependence of these maxima at $z_{\text{max}}$ and the field component $B_\text{x}$ at this position versus $f$. Again, a reasonably short analytical equation could not be found for the $z_{\text{max}}(f)$, but the following expression represents them in the shown range:
\begin{equation}\label{eq:B3}
   z_{\text{max}}(f) \,\approx\, R \sum_{k=1}^{2} A_k \tan^{-1} (k A_0 f) ,
\end{equation}
with $A_0\!=\!0.2833$, $A_1\!=\!0.4242$, and $A_2\!=\!0.3417$. This agrees in the lowest order with Eq.~(\ref{eq:h_max}), which is obtained for $N$=16. 

Finally, the local field at $z_{\text{max}}(f)$ looks like
\begin{equation}\label{eq:B4}
   B_x^{z_{\text{max}}}(f) \,\approx\, \frac{\mu_0 m N}{4\pi R^3} \left(D_0+D_1\frac{D_2}{4f^2+D_2^2} \right),
\end{equation}
with $D_0\!=\!0.115$, $D_1\!=\!4.17$, and $D_2\!=\!3.052$.
Certainly, these properties need to be further explored.

\bibliography{references}

\begin{thebibliography}{34}%
\makeatletter
\providecommand \@ifxundefined [1]{%
 \@ifx{#1\undefined}
}%
\providecommand \@ifnum [1]{%
 \ifnum #1\expandafter \@firstoftwo
 \else \expandafter \@secondoftwo
 \fi
}%
\providecommand \@ifx [1]{%
 \ifx #1\expandafter \@firstoftwo
 \else \expandafter \@secondoftwo
 \fi
}%
\providecommand \natexlab [1]{#1}%
\providecommand \enquote  [1]{``#1''}%
\providecommand \bibnamefont  [1]{#1}%
\providecommand \bibfnamefont [1]{#1}%
\providecommand \citenamefont [1]{#1}%
\providecommand \href@noop [0]{\@secondoftwo}%
\providecommand \href [0]{\begingroup \@sanitize@url \@href}%
\providecommand \@href[1]{\@@startlink{#1}\@@href}%
\providecommand \@@href[1]{\endgroup#1\@@endlink}%
\providecommand \@sanitize@url [0]{\catcode `\\12\catcode `\$12\catcode `\&12\catcode `\#12\catcode `\^12\catcode `\_12\catcode `\%12\relax}%
\providecommand \@@startlink[1]{}%
\providecommand \@@endlink[0]{}%
\providecommand \url  [0]{\begingroup\@sanitize@url \@url }%
\providecommand \@url [1]{\endgroup\@href {#1}{\urlprefix }}%
\providecommand \urlprefix  [0]{URL }%
\providecommand \Eprint [0]{\href }%
\providecommand \doibase [0]{https://doi.org/}%
\providecommand \selectlanguage [0]{\@gobble}%
\providecommand \bibinfo  [0]{\@secondoftwo}%
\providecommand \bibfield  [0]{\@secondoftwo}%
\providecommand \translation [1]{[#1]}%
\providecommand \BibitemOpen [0]{}%
\providecommand \bibitemStop [0]{}%
\providecommand \bibitemNoStop [0]{.\EOS\space}%
\providecommand \EOS [0]{\spacefactor3000\relax}%
\providecommand \BibitemShut  [1]{\csname bibitem#1\endcsname}%
\let\auto@bib@innerbib\@empty
\bibitem [{\citenamefont {Coey}(2002)}]{Coey2002}%
  \BibitemOpen
  \bibfield  {author} {\bibinfo {author} {\bibfnamefont {J.~M.~D.}\ \bibnamefont {Coey}},\ }\bibfield  {title} {\bibinfo {title} {{Permanent magnet applications}},\ }\href {https://doi.org/10.1016/S0304-8853(02)00335-9} {\bibfield  {journal} {\bibinfo  {journal} {Journal of Magnetism and Magnetic Materials}\ }\textbf {\bibinfo {volume} {248}},\ \bibinfo {pages} {441} (\bibinfo {year} {2002})}\BibitemShut {NoStop}%
\bibitem [{\citenamefont {Raich}\ and\ \citenamefont {Blümler}(2004)}]{Raich2004}%
  \BibitemOpen
  \bibfield  {author} {\bibinfo {author} {\bibfnamefont {H.~P.}\ \bibnamefont {Raich}}\ and\ \bibinfo {author} {\bibfnamefont {P.}~\bibnamefont {Blümler}},\ }\bibfield  {title} {\bibinfo {title} {{Design and construction of a dipolar Halbach array with a homogeneous field from identical bar magnets: NMR Mandhalas}},\ }\href {https://doi.org/10.1002/cmr.b.20018} {\bibfield  {journal} {\bibinfo  {journal} {Concepts in Magnetic Resonance Part B - Magnetic Resonance Engineering}\ }\textbf {\bibinfo {volume} {23B}},\ \bibinfo {pages} {16} (\bibinfo {year} {2004})}\BibitemShut {NoStop}%
\bibitem [{\citenamefont {Coey}(473f)}]{Coey_book}%
  \BibitemOpen
  \bibfield  {author} {\bibinfo {author} {\bibfnamefont {J.}~\bibnamefont {Coey}},\ }\href {https://doi.org/10.1017/CBO9780511845000} {\emph {\bibinfo {title} {{Magnetism and Magnetic Materials}}}}\ (\bibinfo  {publisher} {Cambridge University Press},\ \bibinfo {address} {Cambridge},\ \bibinfo {year} {2010, page 473f.})\BibitemShut {NoStop}%
\bibitem [{\citenamefont {Bl\"umler}\ and\ \citenamefont {Casanova}(2015)}]{BluCasa2015}%
  \BibitemOpen
  \bibfield  {author} {\bibinfo {author} {\bibfnamefont {P.}~\bibnamefont {Bl\"umler}}\ and\ \bibinfo {author} {\bibfnamefont {F.}~\bibnamefont {Casanova}},\ }\bibfield  {title} {\bibinfo {title} {{Hardware Developments: Halbach Magnet Arrays}},\ }in\ \href {https://doi.org/10.1039/9781782628095-00133} {\emph {\bibinfo {booktitle} {{Mobile NMR and MRI: Developments and Applications}}}},\ \bibinfo {editor} {edited by\ \bibinfo {editor} {\bibfnamefont {M.}~\bibnamefont {Johns}}, \bibinfo {editor} {\bibfnamefont {E.~O.}\ \bibnamefont {Fridjonsson}}, \bibinfo {editor} {\bibfnamefont {S.}~\bibnamefont {Vogt}},\ and\ \bibinfo {editor} {\bibfnamefont {A.}~\bibnamefont {Haber}}}\ (\bibinfo  {publisher} {Royal Chemical Society},\ \bibinfo {address} {Cambridge},\ \bibinfo {year} {2015})\ \bibinfo {type} {Book section}~\bibinfo {chapter} {4}\BibitemShut {NoStop}%
\bibitem [{\citenamefont {Blümler}(2021)}]{Blümler2021}%
  \BibitemOpen
  \bibfield  {author} {\bibinfo {author} {\bibfnamefont {P.}~\bibnamefont {Blümler}},\ }\bibfield  {title} {\bibinfo {title} {{Magnetic Guiding with Permanent Magnets: Concept, Realization and Applications to Nanoparticles and Cells}},\ }\bibfield  {journal} {\bibinfo  {journal} {CELLS}\ }\textbf {\bibinfo {volume} {10}},\ \href {https://doi.org/10.3390/cells10102708} {10.3390/cells10102708} (\bibinfo {year} {2021})\BibitemShut {NoStop}%
\bibitem [{\citenamefont {Bakenecker}\ \emph {et~al.}(2020)\citenamefont {Bakenecker}, \citenamefont {Schumacher}, \citenamefont {Blümler}, \citenamefont {Gräfe}, \citenamefont {Ahlborg},\ and\ \citenamefont {M~Buzug}}]{Bakenecker2020}%
  \BibitemOpen
  \bibfield  {author} {\bibinfo {author} {\bibfnamefont {A.~C.}\ \bibnamefont {Bakenecker}}, \bibinfo {author} {\bibfnamefont {J.}~\bibnamefont {Schumacher}}, \bibinfo {author} {\bibfnamefont {P.}~\bibnamefont {Blümler}}, \bibinfo {author} {\bibfnamefont {K.}~\bibnamefont {Gräfe}}, \bibinfo {author} {\bibfnamefont {M.}~\bibnamefont {Ahlborg}},\ and\ \bibinfo {author} {\bibfnamefont {T.}~\bibnamefont {M~Buzug}},\ }\bibfield  {title} {\bibinfo {title} {A concept for a magnetic particle imaging scanner with halbach arrays},\ }\href {https://doi.org/10.1088/1361-6560/ab7e7e} {\bibfield  {journal} {\bibinfo  {journal} {Physics in Medicine \& Biology}\ }\textbf {\bibinfo {volume} {65}},\ \bibinfo {pages} {195014} (\bibinfo {year} {2020})}\BibitemShut {NoStop}%
\bibitem [{\citenamefont {Soltner}\ and\ \citenamefont {Blümler}(2023)}]{Soltner2023}%
  \BibitemOpen
  \bibfield  {author} {\bibinfo {author} {\bibfnamefont {H.}~\bibnamefont {Soltner}}\ and\ \bibinfo {author} {\bibfnamefont {P.}~\bibnamefont {Blümler}},\ }\bibfield  {title} {\bibinfo {title} {{Practical Concepts for Design, Construction and Application of Halbach Magnets in Magnetic Resonance}},\ }\href {https://doi.org/10.1007/s00723-023-01602-2} {\bibfield  {journal} {\bibinfo  {journal} {Applied Magnetic Resonance}\ }\textbf {\bibinfo {volume} {54}},\ \bibinfo {pages} {1701} (\bibinfo {year} {2023})}\BibitemShut {NoStop}%
\bibitem [{\citenamefont {Mallinson}(1973)}]{Mallinson1973}%
  \BibitemOpen
  \bibfield  {author} {\bibinfo {author} {\bibfnamefont {J.~C.}\ \bibnamefont {Mallinson}},\ }\bibfield  {title} {\bibinfo {title} {One-sided fluxes - a magnetic curiosity?},\ }\href {https://doi.org/10.1109/TMAG.1973.1067714} {\bibfield  {journal} {\bibinfo  {journal} {IEEE Transactions on Magnetics}\ }\textbf {\bibinfo {volume} {9}},\ \bibinfo {pages} {678} (\bibinfo {year} {1973})}\BibitemShut {NoStop}%
\bibitem [{\citenamefont {Halbach}(1980)}]{Halbach1980}%
  \BibitemOpen
  \bibfield  {author} {\bibinfo {author} {\bibfnamefont {K.}~\bibnamefont {Halbach}},\ }\bibfield  {title} {\bibinfo {title} {{Design of permanent multipole magnets with oriented rare earth cobalt material}},\ }\href {https://doi.org/10.1016/0029-554X(80)90094-4} {\bibfield  {journal} {\bibinfo  {journal} {Nuclear Instruments and Methods}\ }\textbf {\bibinfo {volume} {169}},\ \bibinfo {pages} {1} (\bibinfo {year} {1980})}\BibitemShut {NoStop}%
\bibitem [{Note1()}]{Note1}%
  \BibitemOpen
  \bibinfo {note} {With $N$ cylinders, the center forms a saddle point of order $N$. Field deviations grow $\propto \rho ^N$ from there, as can be interactively explored~\cite {Rehberg2024, Rehberg2025}}\BibitemShut {NoStop}%
\bibitem [{\citenamefont {Jackson}(1999)}]{Jackson1999}%
  \BibitemOpen
  \bibfield  {author} {\bibinfo {author} {\bibfnamefont {J.~D.}\ \bibnamefont {Jackson}},\ }\href {http://cdsweb.cern.ch/record/490457} {\emph {\bibinfo {title} {Classical electrodynamics}}},\ \bibinfo {edition} {3rd}\ ed.\ (\bibinfo  {publisher} {Wiley},\ \bibinfo {address} {New York, {NY}},\ \bibinfo {year} {1999})\BibitemShut {NoStop}%
\bibitem [{\citenamefont {Tewari}\ \emph {et~al.}(2021)\citenamefont {Tewari}, \citenamefont {O'Reilly},\ and\ \citenamefont {Webb}}]{Tewari2021}%
  \BibitemOpen
  \bibfield  {author} {\bibinfo {author} {\bibfnamefont {S.}~\bibnamefont {Tewari}}, \bibinfo {author} {\bibfnamefont {T.}~\bibnamefont {O'Reilly}},\ and\ \bibinfo {author} {\bibfnamefont {A.}~\bibnamefont {Webb}},\ }\bibfield  {title} {\bibinfo {title} {{Improving the field homogeneity of fixed- and variable-diameter discrete Halbach magnet arrays for MRI via optimization of the angular magnetization distribution}},\ }\bibfield  {journal} {\bibinfo  {journal} {Journal of Magnetic Resonance}\ }\textbf {\bibinfo {volume} {324}},\ \href {https://doi.org/10.1016/j.jmr.2021.106923} {10.1016/j.jmr.2021.106923} (\bibinfo {year} {2021})\BibitemShut {NoStop}%
\bibitem [{\citenamefont {Borgers}\ \emph {et~al.}(2018)\citenamefont {Borgers}, \citenamefont {Völkel}, \citenamefont {Schöpf},\ and\ \citenamefont {Reh\-berg}}]{Borgers2018}%
  \BibitemOpen
  \bibfield  {author} {\bibinfo {author} {\bibfnamefont {S.}~\bibnamefont {Borgers}}, \bibinfo {author} {\bibfnamefont {S.}~\bibnamefont {Völkel}}, \bibinfo {author} {\bibfnamefont {W.}~\bibnamefont {Schöpf}},\ and\ \bibinfo {author} {\bibfnamefont {I.}~\bibnamefont {Reh\-berg}},\ }\bibfield  {title} {\bibinfo {title} {Exploring cogging free magnetic gears},\ }\href {https://doi.org/10.1119/1.5029823} {\bibfield  {journal} {\bibinfo  {journal} {American Journal of Physics}\ }\textbf {\bibinfo {volume} {86}},\ \bibinfo {pages} {460} (\bibinfo {year} {2018})},\ \Eprint {https://arxiv.org/abs/https://pubs.aip.org/aapt/ajp/article-pdf/86/6/460/13103287/460\_1\_online.pdf} {https://pubs.aip.org/aapt/ajp/article-pdf/86/6/460/13103287/460\_1\_online.pdf} \BibitemShut {NoStop}%
\bibitem [{\citenamefont {Hartung}\ \emph {et~al.}(2018)\citenamefont {Hartung}, \citenamefont {Sommer}, \citenamefont {V\"olkel}, \citenamefont {Sch\"onke},\ and\ \citenamefont {Reh\-berg}}]{Hartung2018}%
  \BibitemOpen
  \bibfield  {author} {\bibinfo {author} {\bibfnamefont {S.}~\bibnamefont {Hartung}}, \bibinfo {author} {\bibfnamefont {F.}~\bibnamefont {Sommer}}, \bibinfo {author} {\bibfnamefont {S.}~\bibnamefont {V\"olkel}}, \bibinfo {author} {\bibfnamefont {J.}~\bibnamefont {Sch\"onke}},\ and\ \bibinfo {author} {\bibfnamefont {I.}~\bibnamefont {Reh\-berg}},\ }\bibfield  {title} {\bibinfo {title} {Assembly of eight spherical magnets into a dotriacontapole configuration},\ }\href {https://doi.org/10.1103/PhysRevB.98.214424} {\bibfield  {journal} {\bibinfo  {journal} {Phys. Rev. B}\ }\textbf {\bibinfo {volume} {98}},\ \bibinfo {pages} {214424} (\bibinfo {year} {2018})}\BibitemShut {NoStop}%
\bibitem [{\citenamefont {Stamenov}\ and\ \citenamefont {Coey}(2006)}]{Stamenov2006}%
  \BibitemOpen
  \bibfield  {author} {\bibinfo {author} {\bibfnamefont {P.}~\bibnamefont {Stamenov}}\ and\ \bibinfo {author} {\bibfnamefont {J.~M.~D.}\ \bibnamefont {Coey}},\ }\bibfield  {title} {\bibinfo {title} {Vector vibrating-sample magnetometer with permanent magnet flux source},\ }\href {https://doi.org/10.1063/1.2170595} {\bibfield  {journal} {\bibinfo  {journal} {Journal of Applied Physics}\ }\textbf {\bibinfo {volume} {99}},\ \bibinfo {pages} {08D912} (\bibinfo {year} {2006})}\BibitemShut {NoStop}%
\bibitem [{\citenamefont {Soltner}\ and\ \citenamefont {Blümler}(2010)}]{Soltner2010}%
  \BibitemOpen
  \bibfield  {author} {\bibinfo {author} {\bibfnamefont {H.}~\bibnamefont {Soltner}}\ and\ \bibinfo {author} {\bibfnamefont {P.}~\bibnamefont {Blümler}},\ }\bibfield  {title} {\bibinfo {title} {{Dipolar Halbach Magnet Stacks Made from Identically Shaped Permanent Magnets for Magnetic Resonance}},\ }\href {https://doi.org/10.1002/cmr.a.20165} {\bibfield  {journal} {\bibinfo  {journal} {Concepts in Magnetic Resonance}\ }\textbf {\bibinfo {volume} {36A}},\ \bibinfo {pages} {211} (\bibinfo {year} {2010})}\BibitemShut {NoStop}%
\bibitem [{\citenamefont {Bj{\o}rk}(2011)}]{Bjork2011}%
  \BibitemOpen
  \bibfield  {author} {\bibinfo {author} {\bibfnamefont {R.}~\bibnamefont {Bj{\o}rk}},\ }\bibfield  {title} {\bibinfo {title} {{The ideal dimensions of a Halbach cylinder of finite length}},\ }\bibfield  {journal} {\bibinfo  {journal} {Journal of Applied Physics}\ }\textbf {\bibinfo {volume} {109}},\ \href {https://doi.org/10.1063/1.3525646} {10.1063/1.3525646} (\bibinfo {year} {2011})\BibitemShut {NoStop}%
\bibitem [{\citenamefont {Schmidt}\ \emph {et~al.}(2021)\citenamefont {Schmidt}, \citenamefont {Straub}, \citenamefont {Sindersberger}, \citenamefont {Bröckel}, \citenamefont {Monkman},\ and\ \citenamefont {Auernhammer}}]{Schmidt2021}%
  \BibitemOpen
  \bibfield  {author} {\bibinfo {author} {\bibfnamefont {H.}~\bibnamefont {Schmidt}}, \bibinfo {author} {\bibfnamefont {B.~B.}\ \bibnamefont {Straub}}, \bibinfo {author} {\bibfnamefont {D.}~\bibnamefont {Sindersberger}}, \bibinfo {author} {\bibfnamefont {U.}~\bibnamefont {Bröckel}}, \bibinfo {author} {\bibfnamefont {G.~J.}\ \bibnamefont {Monkman}},\ and\ \bibinfo {author} {\bibfnamefont {G.~K.}\ \bibnamefont {Auernhammer}},\ }\bibfield  {title} {\bibinfo {title} {{Collision and separation of nickel particles embedded in a polydimethylsiloxan matrix under a rotating magnetic field: A strong magneto active function.}},\ }\href {https://doi.org/10.1007/s00396-020-04784-4} {\bibfield  {journal} {\bibinfo  {journal} {Colloid and Polymer Science}\ }\textbf {\bibinfo {volume} {299}},\ \bibinfo {pages} {955} (\bibinfo {year} {2021})}\BibitemShut {NoStop}%
\bibitem [{\citenamefont {Blancke~Soares}\ \emph {et~al.}(2022)\citenamefont {Blancke~Soares}, \citenamefont {Gräser}, \citenamefont {Ackers}, \citenamefont {Bakenecker}, \citenamefont {Friedrich}, \citenamefont {Schumacher}, \citenamefont {Lüdtke~Buzug}, \citenamefont {Blümler},\ and\ \citenamefont {Buzug}}]{Soares2022}%
  \BibitemOpen
  \bibfield  {author} {\bibinfo {author} {\bibfnamefont {Y.}~\bibnamefont {Blancke~Soares}}, \bibinfo {author} {\bibfnamefont {M.}~\bibnamefont {Gräser}}, \bibinfo {author} {\bibfnamefont {J.}~\bibnamefont {Ackers}}, \bibinfo {author} {\bibfnamefont {A.}~\bibnamefont {Bakenecker}}, \bibinfo {author} {\bibfnamefont {T.}~\bibnamefont {Friedrich}}, \bibinfo {author} {\bibfnamefont {J.}~\bibnamefont {Schumacher}}, \bibinfo {author} {\bibfnamefont {K.}~\bibnamefont {Lüdtke~Buzug}}, \bibinfo {author} {\bibfnamefont {P.}~\bibnamefont {Blümler}},\ and\ \bibinfo {author} {\bibfnamefont {T.}~\bibnamefont {Buzug}},\ }\bibfield  {title} {\bibinfo {title} {High gradient nested halbach system for steering magnetic particles},\ }\href {https://doi.org/10.18416/ijmpi.2022.2203012} {\bibfield  {journal} {\bibinfo  {journal} {International Journal on Magnetic Particle Imaging}\ ,\ \bibinfo {pages} {Vol 8 No 1 Suppl 1 (2022)}} (\bibinfo {year} {2022})}\BibitemShut {NoStop}%
\bibitem [{Note2()}]{Note2}%
  \BibitemOpen
  \bibinfo {note} {The original equation (6) in Ref.~\cite {Tewari2021} is $\alpha \protect \!=\protect \! 2\varphi + a \cos (2\varphi )$ because a different coordinate system was used (cf.\ their Fig.~1)}\BibitemShut {NoStop}%
\bibitem [{\citenamefont {{International Organization for Standardization}}(2018)}]{Iso9899}%
  \BibitemOpen
  \bibfield  {author} {\bibinfo {author} {\bibnamefont {{International Organization for Standardization}}},\ }\href@noop {} {\bibinfo {title} {Programming languages --- c}},\ \bibinfo {howpublished} {ISO/IEC 9899:2018} (\bibinfo {year} {2018})\BibitemShut {NoStop}%
\bibitem [{\citenamefont {Press}\ \emph {et~al.}(1992)\citenamefont {Press}, \citenamefont {Teukolsky}, \citenamefont {Vetterling},\ and\ \citenamefont {Flannery}}]{Press1992}%
  \BibitemOpen
  \bibfield  {author} {\bibinfo {author} {\bibfnamefont {W.~H.}\ \bibnamefont {Press}}, \bibinfo {author} {\bibfnamefont {S.~A.}\ \bibnamefont {Teukolsky}}, \bibinfo {author} {\bibfnamefont {W.~T.}\ \bibnamefont {Vetterling}},\ and\ \bibinfo {author} {\bibfnamefont {B.~P.}\ \bibnamefont {Flannery}},\ }\href@noop {} {\emph {\bibinfo {title} {Numerical Recipes in C: The Art of Scientific Computing}}},\ \bibinfo {edition} {2nd}\ ed.\ (\bibinfo  {publisher} {Cambridge University Press},\ \bibinfo {year} {1992})\BibitemShut {NoStop}%
\bibitem [{Note3()}]{Note3}%
  \BibitemOpen
  \bibinfo {note} {The unique value for $\alpha _\protect \text {f}$ in programming notation reads $\protect \texttt {mod(atan2(3*sin(2*phi),6*cos(phi)}\wedge \protect \texttt {2-2),2*pi)}$}\BibitemShut {NoStop}%
\bibitem [{\citenamefont {Rehberg}\ and\ \citenamefont {Blümler}(2025)}]{Rehberg2025}%
  \BibitemOpen
  \bibfield  {author} {\bibinfo {author} {\bibfnamefont {I.}~\bibnamefont {Rehberg}}\ and\ \bibinfo {author} {\bibfnamefont {P.}~\bibnamefont {Blümler}},\ }\href {https://doi.org/10.5281/zenodo.15064360} {\bibinfo {title} {Halbach\_two\_point\_oh: Optimize uniform fields with permanent magnet arrays (v1.1.0)}} (\bibinfo {year} {2025})\BibitemShut {NoStop}%
\bibitem [{\citenamefont {Camacho}\ and\ \citenamefont {Sosa}(2013)}]{Camacho2013}%
  \BibitemOpen
  \bibfield  {author} {\bibinfo {author} {\bibfnamefont {J.~M.}\ \bibnamefont {Camacho}}\ and\ \bibinfo {author} {\bibfnamefont {V.}~\bibnamefont {Sosa}},\ }\bibfield  {title} {\bibinfo {title} {Alternative method to calculate the magnetic field of permanent magnets with azimuthal symmetry},\ }\href {https://www.scielo.org.mx/pdf/rmfe/v59n1/v59n1a2.pdf} {\bibfield  {journal} {\bibinfo  {journal} {Revista Mexicana de Física E}\ }\textbf {\bibinfo {volume} {59}},\ \bibinfo {pages} {8} (\bibinfo {year} {2013})}\BibitemShut {NoStop}%
\bibitem [{\citenamefont {Bjørk}\ and\ \citenamefont {d’Aquino}(2023)}]{Bjork2023}%
  \BibitemOpen
  \bibfield  {author} {\bibinfo {author} {\bibfnamefont {R.}~\bibnamefont {Bjørk}}\ and\ \bibinfo {author} {\bibfnamefont {M.}~\bibnamefont {d’Aquino}},\ }\bibfield  {title} {\bibinfo {title} {Accuracy of the analytical demagnetization tensor for various geometries},\ }\href {https://doi.org/https://doi.org/10.1016/j.jmmm.2023.171245} {\bibfield  {journal} {\bibinfo  {journal} {Journal of Magnetism and Magnetic Materials}\ }\textbf {\bibinfo {volume} {587}},\ \bibinfo {pages} {171245} (\bibinfo {year} {2023})}\BibitemShut {NoStop}%
\bibitem [{\citenamefont {Sosa}(2024)}]{Sosa2024}%
  \BibitemOpen
  \bibfield  {author} {\bibinfo {author} {\bibfnamefont {V.}~\bibnamefont {Sosa}},\ }\bibfield  {title} {\bibinfo {title} {Errata of "alternative method to calculate the magnetic field of permanent magnets with azimuthal symmetry"},\ }\href {https://doi.org/10.31349/RevMexFisE.21.020701} {\bibfield  {journal} {\bibinfo  {journal} {Revista Mexicana de Física E}\ }\textbf {\bibinfo {volume} {21}},\ \bibinfo {pages} {020701 1–} (\bibinfo {year} {2024})}\BibitemShut {NoStop}%
\bibitem [{\citenamefont {Tewari}\ and\ \citenamefont {Webb}(2023)}]{Tewari2023}%
  \BibitemOpen
  \bibfield  {author} {\bibinfo {author} {\bibfnamefont {S.}~\bibnamefont {Tewari}}\ and\ \bibinfo {author} {\bibfnamefont {A.}~\bibnamefont {Webb}},\ }\bibfield  {title} {\bibinfo {title} {The permanent magnet hypothesis: an intuitive approach to designing non-circular magnet arrays with high field homogeneity},\ }\bibfield  {journal} {\bibinfo  {journal} {Scientific Reports}\ }\textbf {\bibinfo {volume} {13}},\ \href {https://doi.org/10.1038/s41598-023-29533-9} {10.1038/s41598-023-29533-9} (\bibinfo {year} {2023})\BibitemShut {NoStop}%
\bibitem [{\citenamefont {Insinga}\ \emph {et~al.}(2016)\citenamefont {Insinga}, \citenamefont {Bj{\o}rk}, \citenamefont {Smith},\ and\ \citenamefont {Bahl}}]{Insinga2016}%
  \BibitemOpen
  \bibfield  {author} {\bibinfo {author} {\bibfnamefont {A.~R.}\ \bibnamefont {Insinga}}, \bibinfo {author} {\bibfnamefont {R.}~\bibnamefont {Bj{\o}rk}}, \bibinfo {author} {\bibfnamefont {A.}~\bibnamefont {Smith}},\ and\ \bibinfo {author} {\bibfnamefont {C.~R.~H.}\ \bibnamefont {Bahl}},\ }\bibfield  {title} {\bibinfo {title} {Globally optimal segmentation of permanent-magnet systems},\ }\bibfield  {journal} {\bibinfo  {journal} {Phys. Rev. Appl.}\ }\textbf {\bibinfo {volume} {5}},\ \href {https://doi.org/10.1103/PhysRevApplied.5.064014} {10.1103/PhysRevApplied.5.064014} (\bibinfo {year} {2016})\BibitemShut {NoStop}%
\bibitem [{\citenamefont {Insinga}\ \emph {et~al.}(2019)\citenamefont {Insinga}, \citenamefont {Smith}, \citenamefont {Bahl}, \citenamefont {Nielsen},\ and\ \citenamefont {Bj{\o}rk}}]{Insinga2019}%
  \BibitemOpen
  \bibfield  {author} {\bibinfo {author} {\bibfnamefont {A.~R.}\ \bibnamefont {Insinga}}, \bibinfo {author} {\bibfnamefont {A.}~\bibnamefont {Smith}}, \bibinfo {author} {\bibfnamefont {C.~R.~H.}\ \bibnamefont {Bahl}}, \bibinfo {author} {\bibfnamefont {K.~K.}\ \bibnamefont {Nielsen}},\ and\ \bibinfo {author} {\bibfnamefont {R.}~\bibnamefont {Bj{\o}rk}},\ }\bibfield  {title} {\bibinfo {title} {Optimal segmentation of three-dimensional permanent-magnet assemblies},\ }\bibfield  {journal} {\bibinfo  {journal} {Phys. Rev. Appl}\ }\textbf {\bibinfo {volume} {12}},\ \href {https://doi.org/10.1103/PhysRevApplied.12.064034} {10.1103/PhysRevApplied.12.064034} (\bibinfo {year} {2019})\BibitemShut {NoStop}%
\bibitem [{\citenamefont {Rehberg}(2024)}]{Rehberg2024}%
  \BibitemOpen
  \bibfield  {author} {\bibinfo {author} {\bibfnamefont {I.}~\bibnamefont {Rehberg}},\ }\href {https://doi.org/10.5281/zenodo.11214206} {\bibinfo {title} {{From Concentric Halbach Rings to Tetraplex Cuts – Examine 540 Dipole Clusters with a Single Python Animation}}} (\bibinfo {year} {2024})\BibitemShut {NoStop}%
\bibitem [{\citenamefont {Casanova}\ \emph {et~al.}(2011)\citenamefont {Casanova}, \citenamefont {Perlo},\ and\ \citenamefont {Bl\"umich}}]{Casanova2011}%
  \BibitemOpen
  \bibfield  {author} {\bibinfo {author} {\bibfnamefont {F.}~\bibnamefont {Casanova}}, \bibinfo {author} {\bibfnamefont {J.}~\bibnamefont {Perlo}},\ and\ \bibinfo {author} {\bibfnamefont {B.}~\bibnamefont {Bl\"umich}},\ }\href {https://doi.org/10.1007/978-3-642-16307-4} {\emph {\bibinfo {title} {Single-Sided NMR}}}\ (\bibinfo  {publisher} {Springer},\ \bibinfo {address} {Heidelberg},\ \bibinfo {year} {2011})\BibitemShut {NoStop}%
\bibitem [{\citenamefont {Blümler}\ and\ \citenamefont {Casanova}(2015)}]{Casanova2016}%
  \BibitemOpen
  \bibfield  {author} {\bibinfo {author} {\bibfnamefont {P.}~\bibnamefont {Blümler}}\ and\ \bibinfo {author} {\bibfnamefont {F.}~\bibnamefont {Casanova}},\ }\bibfield  {title} {\bibinfo {title} {Hardware developments: Single-sided magnets},\ }in\ \href {https://doi.org/10.1039/9781782628095-00110} {\emph {\bibinfo {booktitle} {Mobile NMR and MRI: Developments and Applications}}}\ (\bibinfo  {publisher} {The Royal Society of Chemistry},\ \bibinfo {year} {2015})\BibitemShut {NoStop}%
\bibitem [{\citenamefont {Hobson}\ \emph {et~al.}(2024)\citenamefont {Hobson}, \citenamefont {Morley}, \citenamefont {Davis}, \citenamefont {Smith},\ and\ \citenamefont {Fromhold}}]{Hobson2024}%
  \BibitemOpen
  \bibfield  {author} {\bibinfo {author} {\bibfnamefont {P.~J.}\ \bibnamefont {Hobson}}, \bibinfo {author} {\bibfnamefont {C.}~\bibnamefont {Morley}}, \bibinfo {author} {\bibfnamefont {A.}~\bibnamefont {Davis}}, \bibinfo {author} {\bibfnamefont {T.}~\bibnamefont {Smith}},\ and\ \bibinfo {author} {\bibfnamefont {M.}~\bibnamefont {Fromhold}},\ }\bibfield  {title} {\bibinfo {title} {Target-field design of surface permanent magnets},\ }\bibfield  {journal} {\bibinfo  {journal} {Phys. Rev. Appl.}\ }\textbf {\bibinfo {volume} {22}},\ \href {https://doi.org/10.1103/PhysRevApplied.22.034015} {10.1103/PhysRevApplied.22.034015} (\bibinfo {year} {2024})\BibitemShut {NoStop}%
\end{thebibliography}%

\end{document}